\documentclass[useAMS,usenatbib,usegraphicx,onecolumn]{mn2e}

\usepackage{amssymb,amsfonts,amsmath,amstext,amsgen,amsopn,amsxtra,indentfirst,lscape,times}

\def\yr{{\rm\,yr}}
\def\km{{\rm\,km}}
\def\kms{\,\hbox{km s}^{-1}}
\def\kpc{{\rm\,kpc}}
\def\au{{\rm\,AU}}
\def\Gyr{{\rm\,Gyr}}
\def\gmcm{\,\hbox{g cm}^{-3}}
\def\gm{{\rm\,g}}
\def\be{\begin{equation}}
\def\ee{\end{equation}}
\def\ffrac#1#2{{\textstyle\frac{#1}{#2}}}
\def\invisible#1{}
\def\half{{\textstyle{\frac{1}{2}}}}
\def\apj{ApJ}
\def\apjl{ApJL}
\def\apjs{ApJS}
\def\mnras{MNRAS}
\def\aap{A\&A}
\def\araa{ARA\&A}
\def\jgr{JGR}
\def\planss{Planet. Space Sci.}

\title[Long-Lived Planetesimal Discs]{Long-Lived Planetesimal Discs}
\author[K. Heng and S. Tremaine]{Kevin Heng\thanks{E-mail: heng@ias.edu (KH); tremaine@ias.edu (ST)} and Scott Tremaine\footnotemark[1]\\
Institute for Advanced Study, School of Natural Sciences, Einstein Drive, Princeton, NJ 08540, U.S.A.}
\begin{document}

\date{Accepted 2009 Sept 14.  Received 2009 Sept 7; in original form 2009 June 10}

\pagerange{\pageref{firstpage}--\pageref{lastpage}} \pubyear{2009}

\maketitle

\label{firstpage}

\begin{abstract}
  We investigate the survival of planetesimal discs over Gyr timescales, using
  a unified approach that is applicable to all Keplerian discs of solid bodies
  -- dust grains, asteroids, planets, etc.  Planetesimal discs can be
  characterized locally by four parameters: surface density, semi-major axis,
  planetesimal size and planetesimal radial velocity dispersion.  Any
  planetesimal disc must have survived all dynamical processes, including
  gravitational instability, dynamical chaos, gravitational scattering,
  physical collisions, and radiation forces, that would lead to significant
  evolution over its lifetime. These processes lead to a rich set of
  constraints that strongly restrict the possible properties of long-lived
  discs. Within this framework, we also discuss the detection of planetesimal
  discs using radial velocity measurements, transits, microlensing, and the
  infrared emission from the planetesimals themselves or from dust
  generated by planetesimal collisions.
\end{abstract}

\begin{keywords}
  planets and satellites: formation -- Kuiper Belt -- minor planets, asteroids
  -- gravitational lensing -- Solar system: formation -- stars: formation
\end{keywords}

\section{Introduction}

Terrestrial planets and the cores of giant planets are generally believed to
have formed hierarchically: small solid bodies (`planetesimals') condense
from the gaseous circumstellar disc \citep{jo07}, collide repeatedly, and
accumulate into larger and larger assemblies \citep{saf72,gls04a,rei07}.
Several aspects of this complex process remain obscure, in particular (but not
limited to) the formation of planetesimals from dust grains \citep{bw08}, how
to grow Uranus and Neptune in the short time available before the gaseous disc
is dissipated \citep{gls04a,gls04b}, the origins of the large eccentricities
of the extrasolar planets \citep{tz04}, the role of planetary migration
\citep{gt80,pt06}, how the residual planetesimals were cleaned out of the
solar system \citep{gls04b}, and why the solar system is so different from
known extrasolar planetary systems \citep{bee04}.

Given these large gaps in our understanding, it is worthwhile to investigate
not just the difficult question of how planets form but also the simpler
question of whether they can survive once formed. Planetary systems in the
Galaxy have presumably been formed at a more-or-less constant rate, so it is
reasonable to assume that most of today's planetary systems are at least
several Gyr old. They must therefore have survived all dynamical
processes -- gravitational instabilities, collisions, viscous stirring or
two-body relaxation, etc. -- that would lead to a substantial change in their
properties on timescales less than about 3 Gyr. Understanding what long-lived
planetary systems are possible should help us to understand to what extent the
properties of actual planets are shaped by the formation process as opposed to
evolution (`nature versus nurture') and may guide observers in
searching for novel types of planetary systems.

Orbiting solid bodies are often called `planetesimals', `planetary
embryos', or `planets' depending on their mass, but for simplicity we shall
use the term `planetesimal' to describe any body, whether solid like a
terrestrial planet or gas-dominated like a giant planet, that is large enough
so that gas drag and radiation effects (Poynting-Robertson drag, radiation
pressure, Yarkovsky effect, etc.) are negligible.

We shall find it useful to classify discs as `hot' or `cold' depending on
whether or not the planetesimal orbits cross. Within the solar system, the
planets form a cold system (except for Pluto) while the asteroid and Kuiper
belts are hot. Among hot planetesimal discs, an important special case is the
`collision-limited' disc, in which the collision time between planetesimals is
equal to the age of the disc. Collision-limited discs are likely to arise from
discs in which there is a distribution of planetesimal sizes: smaller
planetesimals have shorter collision times and therefore are destroyed first,
so the dominant planetesimal population (by mass) always has a collision time
that is roughly equal to the disc age.  We shall also use the term `warm' to
describe discs in which the planetesimal orbits cross, but the impact
velocities are so low that the cumulative effect of collisions does not
substantially damage the planetesimals (\S\ref{subsect:dynamics_collisions}).

The most important observational signature of many planetesimal discs arises
from dust formed in recent planetesimal collisions. These `debris discs'
were first detected from the thermal emission of the dust, which creates an
infrared (IR) excess in the spectral energy distribution of the otherwise
normal stars that they surround \citep{au84}. Spatially resolved debris discs
are sometimes also visible from their scattered light.  The IR excess
(bolometric) luminosity relative to the luminosity from the parent star
depends on the distance and spectral type of the host star as well as the disc
radius, but is typically $\gtrsim 10^{-5}$ (see \citealt{zu01} and
\citealt{wy08} for reviews).  The asteroid and Kuiper belts in our own solar
system can be thought of as debris discs, although they would not be
detectable around other stars with current technology since the bolometric IR
excess is $\sim 10^{-7}$ for both belts.

Debris discs have been detected around stars with a wide range of spectral
types (A to M) and ages ($\sim 10^7$ to $10^{10}$ yr)\footnote{IR excesses
  characteristic of debris discs have also been detected around white dwarfs,
  and they appear to be strongly correlated with metal contamination in the
  white-dwarf photosphere, presumably arising from accreted planetesimals
  \citep{far09}.}.  The dust masses inferred from these observations are
$10^{-3} \lesssim M_{\rm dust}/M_\oplus \lesssim 1$ (Fig. 3 of
\citealt{wy08}), although this result depends on the assumed size distribution
of the dust.  The term `debris' emphasizes that the lifetime of the dust
grains from grain-grain collisions, Poynting--Robertson drag, or radiation
pressure is robustly and considerably less than the stellar age, so the dust
cannot be primordial and must be continuously regenerated, presumably by
ongoing planetesimal collisions.

We stress that the discs we consider in this paper are far more general than
debris discs: they include hot discs in which the rate of dust generation may
be undetectably small, cold discs in which there are no collisions, planetary
systems, asteroid belts, planetary rings, etc.

We begin by constructing a simple model for a planetesimal disc in
\S\ref{sec:simple}.  In \S\ref{sec:mono}, we describe the dynamical processes
that act on planetesimal discs.  Collision-limited discs, which are a special
subset of hot discs, are described in \S\ref{sec:col}.  Non-gravitational
forces on dust are briefly reviewed in \S\ref{sec:dust}.  The properties of
long-lived discs are discussed in \S\ref{sect:longliveddiscs}, where we also
study six sample discs (Table \ref{tab:disc}).  We discuss
possible techniques for detecting and studying planetesimal discs in
\S\ref{sect:detection}, and we summarize and discuss our results in
\S\ref{sec:disc}. 

\begin{table*}
\centering
\begin{minipage}{140mm}
  \caption{Sample planetesimal discs}
  \label{tab:disc}
  \begin{tabular}{lcccccc}
\hline \hline
\multicolumn{1}{c}{} & \multicolumn{1}{c}{A} & \multicolumn{1}{c}{B} & \multicolumn{1}{c}{C} & \multicolumn{1}{c}{D} & \multicolumn{1}{c}{E} & \multicolumn{1}{c}{F}\\
\hline
\vspace{0.06in}
$a$                      
    & $1\au$      & $1\au$    & $10\au$     & $10\au$   & $100\au$ & $100\au$ \\
\vspace{0.06in}
$\mu \equiv \Sigma a^2/M_\odot$
    & $10^{-4}$    & $10^{-6}$   &  $10^{-4}$   & $10^{-6}$  &  $10^{-4}$   & $10^{-6}$ \\
\vspace{0.06in}
$M_{\rm disc}=\pi f_m \Sigma a^2$ 
    & $\simeq 100M_\oplus$ & $\simeq 1M_\oplus$ & $\simeq 100M_\oplus$ & $\simeq 1M_\oplus$ &
    $\simeq 100M_\oplus$ & $\simeq 1M_\oplus$ \\
\vspace{0.06in}
$m_{\rm min}^{\rm cold}$ 
    & 60 $M_\oplus$ & 0.1 $M_\oplus$ & 60 $M_\oplus$ & 0.1 $M_\oplus$ & 60 $M_\oplus$ & 0.1 $M_\oplus$ \\
\vspace{0.06in}
$N_{\rm max}^{\rm cold}$
    & 1--2 & 12 & 1--2 & 12 & 1--2 & 12 \\
\vspace{0.06in}
$m_{\rm min}^{\rm hot}$ 
    & -- & -- & -- & $10^{24}$ g & $4 \times 10^{19}$ g & $4 \times 10^{13}$ g \\
\vspace{0.06in}
$N_{\rm max}^{\rm hot}$
    & -- & -- & -- & $5 \times 10^3$ & $2 \times 10^{10}$ & $2 \times 10^{14}$ \\
\vspace{0.06in}
$m_{\rm max}^{\rm warm}$ 
    & -- & -- & -- & $3 \times 10^{25}$ g & $10^{21}$ g & $10^{15}$ g \\
\vspace{0.06in}
$N_{\rm min}^{\rm warm}$
    & --& -- & -- & 200 & $5 \times 10^8$ & $5 \times 10^{12}$ \\
\hline
\end{tabular}

Note: $m_{\rm min}^{\rm cold}$ (equation [\ref{eq:mmincold}]), $m_{\rm min}^{\rm hot}$ (equation [\ref{eq:hotdiscc}]) are the minimum planetesimal mass for cold and hot discs, respectively; $m_{\rm max}^{\rm warm}$ is the maximum mass for warm discs (equation [\ref{eq:warmdisc_mmax}]).  $N_{\rm max}^{\rm cold}$ (equation [\ref{eq:nmaxa}]) and $N_{\rm max}^{\rm hot}$ (equation [\ref{eq:hotdiscc}]) are the maximum number of planetesimals per octave for cold and hot discs, respectively; $N_{\rm min}^{\rm warm}$ (equation [\ref{eq:warmdisc_mmax}]) is the minimum number for warm discs.
\end{minipage}
\end{table*}

\section{A simple model for a planetesimal disc}
\label{sec:simple}

We consider a system of planetesimals orbiting a solar-type star of mass $M_\odot$. The extension to other types of stars is straightforward, but at this
stage adds excessive complication. We shall focus on discs with an age
$t_0=3\Gyr$ since these are likely to be more common than younger discs. We
assume that the discs are gas-free, which is consistent with observations for
most discs older than a few Myr (see Fig. 2 of \citealt{wy08}).

The surface density of the disc at semi-major axis $a$ is written 
$\Sigma(a)$; more precisely, the mass with semi-major axes in the
range $[a,a+da]$ is $dM_{\rm disc}=2\pi\Sigma a\,da$. The orbital period is $2\pi/\Omega$, where
\be
\Omega=\sqrt{\frac{GM_\odot}{a^3}}.
\ee

In most cases, we shall assume that this material is collected in identical
spherical bodies of density $\rho_p$, radius $r$, and mass
$m=4\pi\rho_p r^3/3$ -- a monodisperse planetesimal system (see the end of
\S\ref{subsect:coll_limited} for further discussion of this approximation). The
densities of the planets in the solar system range 
from $5.5\gmcm$ (Earth) to $0.7\gmcm$ (Saturn); in general gas-giant planets
have smaller densities than terrestrial planets, but we shall sacrifice
accuracy for simplicity and assume that all planetesimals have
$\rho_p=3\gmcm$. The surface number density of planetesimals is 
\be 
{\cal N}=\frac{\Sigma}{m}.
\ee

We assume that the eccentricities $e$ and inclinations $i$ of the
planetesimals follow a Rayleigh distribution, 
\be 
  d^2n \propto ei\exp\left[-\left(e/e_0\right)^2 - \left(i/i_0\right)^2 \right]di\,de,
\label{eq:rayleigh}
\ee
where $e_0$ and $i_0$ are the root mean square (rms) eccentricity and inclination. The density of
the disc in the direction normal to its symmetry plane (which we call the
$z$-direction) is given by 
\be
  d{\cal N}=n(z)\,dz=n_0\exp(-\half z^2/h^2)\,dz,
\label{eq:nz}
\ee
where $h\equiv ai_0/\surd{2}$ is the rms $z$-coordinate. The midplane number 
density $n_0$ is related to the surface number density by   
\be
n_0=\frac{\cal N}{\sqrt{2\pi}h}=\frac{\Sigma}{\sqrt{\pi}mai_0}.
\label{eq:rhosig}
\ee
The radial velocity dispersion is 
\be
\sigma_r=\frac{\Omega ae_0}{\surd{2}}. 
\label{eq:vdisp}
\ee
The ratio $i_0/e_0$ can in principle have a wide range of values depending on
the dynamical history of the disc, but in a variety of theoretical models and
observed astrophysical discs $i_0/e_0 \simeq 0.5$ \citep{dt93,si00} so we shall
adopt this value throughout the paper (see also discussion following equation
[\ref{eq:tcollvv}]). 

With these assumptions, the local properties of the planetesimal disc are
specified by four parameters: the semi-major axis $a$, the surface density
$\Sigma$, the planetesimal radius $r$ or mass $m$, and the rms eccentricity
$e_0$.  (We take the planetesimal density $\rho_p$, the disc age $t_0$, and the
stellar mass $M_\odot$ to be fixed.) Collision-limited discs are specified by
three parameters, since the requirement that the collision time equals the age
provides one constraint on the four parameters.

We shall find it useful to express many of our results in terms of the
following three dimensionless parameters:
\begin{eqnarray}
\mu & \equiv & \frac{\Sigma a^2}{M_\odot},\nonumber \\
\nu & \equiv & \frac{\rho_p a^3}{M_\odot}= 5.0\times 10^9 \left(\frac{\rho_p}
{3\gmcm}\right) \left(\frac{a}{10\au}\right)^3, \nonumber \\
\tau & \equiv & \frac{\Sigma a^2}{M_\odot}\Omega t_0 = 6.0\times 10^4 \left(\frac{\mu}{10^{-4}} \right) \left( \frac{a}{10\au}\right)^{-3/2} \left(\frac{t_0}{3 \Gyr} \right).
\label{eq:dimen}
\end{eqnarray}
The parameter $\mu$ is a dimensionless mass, of order the ratio of the disc
mass to the stellar mass. The parameter $\nu$ is a dimensionless density,
which is almost always quite large for planetesimal discs, so we may 
assume $\nu\gg 1$ when necessary.  The parameter $\tau$ is a
dimensionless age, scaled by $\Omega$ and by $\Sigma a^2/M_\odot$ since most
evolutionary processes run slower if the orbital time is longer or the disc
contains less mass.

We parametrize discs in terms of surface density $\Sigma$ and semi-major axis
$a$ since these are likely to vary much less than planetesimal mass $m$ or rms
eccentricity $e_0$ during the planet formation process.  As points of
reference, we shall refer to a set of six discs (Table \ref{tab:disc}) with
semi-major axes $a=1\au$, $a=10\au$ and $a=100\au$ (a typical radius for
spatially resolved debris discs), and with
dimensionless mass $\mu=10^{-4}$ and $10^{-6}$. The former mass corresponds to
$M_{\rm disc} \sim \pi \Sigma a^2 = \pi\mu M_\odot \simeq 0.3 M_{\rm
  Jupiter} \simeq 100M_\oplus$, about the solid mass needed to form the giant planets
and comets in the solar system \citep{gls04b}, while the latter mass corresponds to
$M_{\rm disc} \simeq M_\oplus$.

\section{Dynamical processes in the planetesimal disc}
\label{sec:mono}

\begin{table*}
\centering
\begin{minipage}{140mm}
\caption{Summary of dimensionless numbers}
\label{tab:nums}
\begin{tabular}{lccc}
\hline\hline
\multicolumn{1}{c}{Symbol} & \multicolumn{1}{c}{Assumed value} & \multicolumn{1}{c}{Defining equation and/or section}\\
\hline
\vspace{2pt}
$i_0/e_0$ & 0.5 & (\ref{eq:rayleigh})\\
$f_e$ & 0.5 & (\ref{eq:thin})\\
$f_m$ & 1 & (\ref{eq:fm})\\
$f_d$ & 0.3 & (\ref{eq:discmass})\\
$f_c$ & 1 & (\ref{eq:hot})\\
$f_Q$ & 1 & (\ref{eq:q}), (\ref{eq:tt}) \\
$f_1$ & 0.690 & (\ref{eq:tcolla}), Appendix \ref{append:dt93}\\
$f_2$ & 1.521 & (\ref{eq:tcolla}), Appendix \ref{append:dt93}\\
$f_3$ & 0.28 & (\ref{eq:trelax})\\
$f_4$ & 22.67 & (\ref{eq:tcollshear}), Appendix \ref{append:dt93}\\
$f_5$ & 12.94 & (\ref{eq:tcollshear}), Appendix \ref{append:dt93}\\
$f_6$ & 1 & (\ref{eq:spread})\\
$f_7$ & 0.46 & (\ref{eq:abc}) \\
$\mu$ & various & (\ref{eq:dimen})\\
$\nu$ & various & (\ref{eq:dimen})\\
$\tau$ & various & (\ref{eq:dimen})\\
$a$ & $-0.13$ & (\ref{eq:qast})\\
$b$ & 0.44 & (\ref{eq:qast})\\
\hline
\end{tabular}
\end{minipage}
\end{table*}

There are a few general criteria that planetesimal discs must satisfy.  Firstly,
we require that the planetesimal eccentricities and inclinations are not too
large, which can be interpreted as the condition that most of
the planetesimals are bound ($e<1$), or that the disc is thin ($h < a$),
or that the radial velocity dispersion is less than the circular speed
($\sigma_r < \Omega a$). We shall write this criterion as 
\be
e_0 \lesssim f_e\quad\hbox{or}\quad \frac{h}{a}\lesssim 0.35f_e\quad \hbox{or}\quad
\frac{\sigma_r}{\Omega a} \lesssim \frac{f_e}{\surd{2}}.
\label{eq:thin}
\ee
In this paper, we adopt $f_e=0.5$. (See Table \ref{tab:nums} for a summary of the dimensionless numbers
used in our study.)

Naturally, non-trivial planetesimal discs must contain more than one body.
Assuming that the surface density does not vary strongly with radius and the
disc is not too extended, the disc mass may be written
\be
M_{\rm disc}=\pi f_m \Sigma a^2
\label{eq:fm}
\ee
with $f_m$ of order unity, so the criterion that the number of
planetesimals $N= M_{\rm disc}/m > 1$ becomes
\be
m < \pi f_m\Sigma a^2 \quad\hbox{or}\quad \pi f_m{\cal N}a^2 > 1.
\label{eq:number}
\ee
In this paper we adopt $f_m=1$, which corresponds to a disc with a radial width of about $0.5a$.

We shall also require that the total disc mass is small compared to the stellar
mass, 
\be
\frac{M_{\rm disc}}{M_\odot}< f_d,
\label{eq:discmass}
\ee
where we adopt $f_d=0.3$. 

The total cross-sectional area of the disc is $\pi r^2N$; thus, in the absence
of mutual shadowing the geometrical optical depth of the disc as seen from the
host star is 
\be
\tau_p= \frac{N r^2}{4a^2} = \frac{f_m}{4} {\cal N} \pi r^2.
\label{eq:taupdef}
\ee
Note that this is differs by a factor of order unity from the normal
geometrical optical depth of the ring,  ${\cal N} \pi r^2$.

\subsection{Cold discs}
\label{subsect:colddiscs}

A `cold' disc is one in which planetesimal orbits do not cross. A planetesimal in an
orbit with eccentricity $e$ has a total radial excursion of $2ae$. The typical
radial separation between planetesimals is 
\be
  \Delta a=\frac{m}{2\pi \Sigma a}.
\label{eq:da}
\ee
Thus most orbits do not cross if the rms eccentricity is
\be 
e_0 < e_{\rm hot}\equiv f_c\frac{m}{4\pi \Sigma a^2}, 
\label{eq:hot}
\ee 
and we shall choose $f_c=1$, at which point just over half of the particles
cross if their semi-major axes have a Poisson distribution.  (A refinement of
the preceding condition is to include the radius of the planetesimal in the
crossing condition, that is, $2(ae+r)<\Delta a$, but in the cases of interest
to us this correction is unimportant.) Equivalently, one can describe cold
discs as those for which $\sigma_r < \sigma_{\rm hot}$, where

\be 
\sigma_{\rm hot} \equiv \frac{f_c}{2^{5/2}\pi}\frac{m\Omega}{a\Sigma} = \frac{f_c}{3\sqrt{2}}\frac{\sqrt{GM_\odot} \rho_p r^3}{a^{5/2} \Sigma}. 
\label{eq:shot} 
\ee

\subsubsection{Gravitational stability}

Gravitational instability in cold discs is associated with dynamical chaos,
which leads to growth in the eccentricities and inclinations of the
planetesimals. The growth rate can be extremely slow -- for example, tens of
Gyr for Mercury in the current solar system configuration \citep{las08}. No
rigorous analytical formulae for the rate of chaotic evolution in cold discs
are available. However, N-body experiments suggest that cold discs can survive
for millions of orbits if the separation (\ref{eq:da}) is typically a few
times larger than the Hill radius
\be
r_{\rm H} \equiv a \left(\frac{m}{3M_\odot}\right)^{1/3}
\label{eq:hillrad}
\ee
Thus, stability requires\invisible{actual value 9.0930}
\be
\frac{\Delta a}{a} > k\left(\frac{m}{3M_\odot}\right)^{1/3}\quad \hbox{or} \quad 
\frac{m}{M_\odot} > 9.1 \left(k \mu \right)^{3/2}.
\label{eq:hillstab}
\ee
\cite{cwb96} find $k=7$--11 from integrations of planetesimals with masses
between $10^{-5}$ and $10^{-7}M_\odot$ lasting $\sim 10^7\yr$.\footnote{Note
that \cite{cwb96} define the Hill radius as $(2m/3M_\odot)^{1/3}$, a factor
$2^{1/3}$ larger than our definition.} They also find that the
factor $k$ varies slowly with planetesimal mass, so a stability
criterion that fits their results more accurately is
\be
\frac{\Delta a}{a} > 4\left(\frac{m}{M_\odot}\right)^{0.3}\quad \hbox{or}\quad 
\frac{m}{M_\odot} > 100\mu^{1.43}.
\label{eq:hillstabb}
\ee
This result depends on the duration of the integrations, but only weakly:
increasing the duration by a factor of ten typically increases the minimum
stable separation by about one Hill radius. 

The criterion (\ref{eq:hillstabb}) was derived from simulations with
planetesimals of mass $10^{-5}$--$10^{-7}M_\odot$; the extrapolation to larger
planets is somewhat uncertain but probably not a major source of error. For
Jupiter-mass planets the formula predicts $\Delta a/a > 0.5$, corresponding to
$k \simeq 7$ in the notation of equation (\ref{eq:hillstab}). For comparison,
\cite{jt08} estimate $k \simeq 12$--14 from orbit integrations of planets with
masses between 0.1 and 10 Jupiter masses; their larger value of $k$ probably
arises because they used a range of masses rather than a single common mass
for the planets. In any event, equation (\ref{eq:hillstabb}) should be correct
to within a factor of two or so.

When the planetesimals in a cold disc have non-zero eccentricities, larger
separations are required for stability  \cite[e.g.,][]{ykm99}; however, we shall
not include this refinement since the detectability of cold discs does not
depend on the rms eccentricity (equation [\ref{eq:hot}]). 

\subsection{Hot discs}

A hot disc is one in which planetesimal orbits cross. In hot discs, the rms
eccentricity or radial velocity dispersion exceeds $e_{\rm hot}$ or
$\sigma_{\rm hot}$ respectively (equations [\ref{eq:hot}] and
[\ref{eq:shot}]).  As we discuss below (equation [\ref{eq:hotdiscs_type}]),
`warm' discs may also be defined, in which the orbits cross but the
collisions do not substantially damage the planetesimals over the lifetime of
the disc.

\subsubsection{Gravitational stability}
\label{subsect:grav_stab}

The disc must be gravitationally stable to axisymmetric
perturbations \citep{too64}. The Toomre stability criterion is derived from
the WKB dispersion relation for axisymmetric density waves. For a fluid disc
in a Keplerian potential this may be written as \citep{bt08}
\be
\frac{\omega^2}{\Omega^2}=1-\frac{\lambda_{\rm crit}}{\lambda}+\frac{Q^2}{4}
\frac{\lambda_{\rm     crit}^2}{\lambda^2},
\label{eq:wkb}
\ee
where $\omega$ is the frequency, $\lambda$ is the radial wavelength, 
and 
\be
\begin{split}
&Q \equiv \frac{\sigma_r \Omega}{\pi G \Sigma}=\frac{e_0}{\sqrt{2}\pi\mu},\\
&\lambda_{\rm crit}=\frac{4\pi^2 G\Sigma}{\Omega^2} = 4\pi^2 a \mu.\\
\end{split}
\ee
This dispersion relation was derived for a barotropic fluid 
whereas the planetesimal disc more closely approximates a collisionless fluid,
but for our purposes the results should be accurate enough.

The disc is stable to perturbations at a given wavelength $\lambda$ if
$\omega^2>0$, which requires 
\be
Q^2 > 4\frac{\lambda}{\lambda_{\rm crit}}\left(1-\frac{\lambda}{\lambda_{\rm
      crit}}\right). 
\label{eq:q}
\ee
The maximum of the right side occurs at $\lambda=\lambda_{\rm crit}/2$ and
equals unity, so the disc is stable to perturbations of all wavelengths if
$Q>1$. However, this result needs to be modified if the number of planetesimals
is so small that the disc cannot be approximated as a continuous
fluid. The typical separation between planetesimals is given by equation (\ref{eq:da}),
so the number of planetesimals in one wavelength $\lambda$ is
$N_\lambda=\lambda/\Delta a=2\pi a\Sigma \lambda/m$.  The continuum
approximation should be valid if $N_\lambda \gg 1$ or $\lambda >
\lambda_c\equiv f_Q m/(2\pi a\Sigma)$ with $f_Q$ of order unity. We adopt $f_Q=1$
when a choice is necessary. Then a necessary requirement for stability is that
equation (\ref{eq:q}) is satisfied for all $\lambda > \lambda_c$, or
\be
Q^2 > \left\{\begin{array}{ll} 1 & \mbox{if $\lambda_c < \lambda_{\rm crit}/2$},
    \\
                   4(\lambda_c/\lambda_{\rm crit})- 4(\lambda_c/\lambda_{\rm
                     crit})^2 & \mbox{if $\lambda_c > \lambda_{\rm crit}/2$}.
             \end{array}\right.
\ee
This may be rewritten as
\begin{eqnarray}
e_0 & > & \sqrt{2}\pi \mu, \qquad \mbox{if} \quad \frac{m}{M_\odot}<\frac{4\pi^3\mu^2}{f_Q} , \nonumber \\
e_0 & > &  \sqrt{\frac{f_Q m}{\pi M_\odot} \left[1-\frac{f_Q m}{ \left(2\pi\right)^3 M_\odot \mu^2} \right]}, \quad \mbox{otherwise}; 
\label{eq:tt} 
\end{eqnarray}
there is no constraint if the square root in the second equation is negative. 

\begin{figure}
\begin{center}
\includegraphics[width = 0.8\columnwidth]{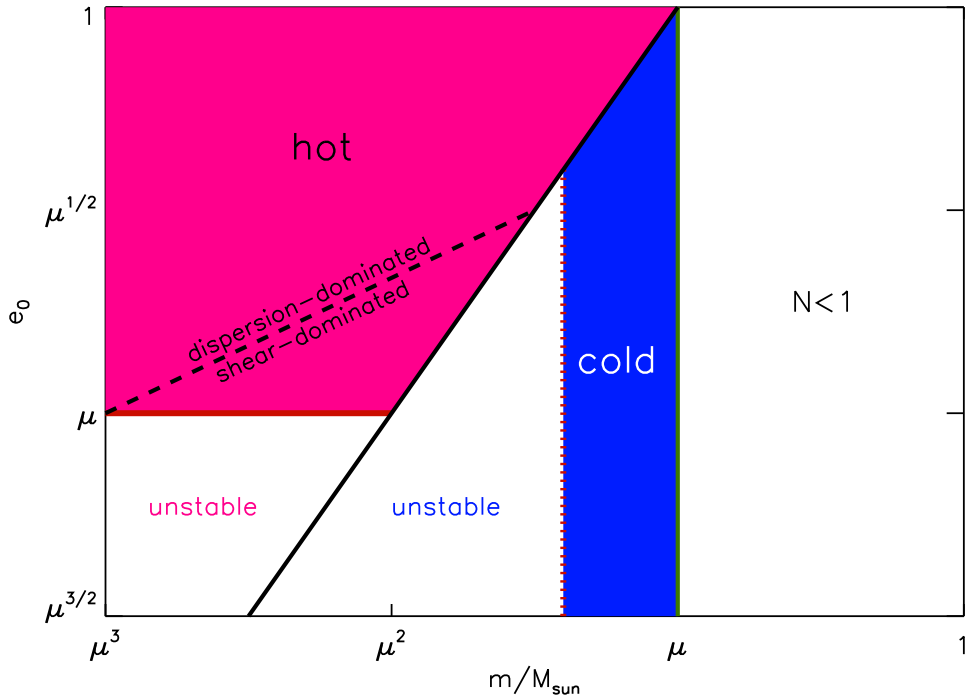}
\end{center}
\caption{Stability and other properties of planetesimal discs, as a function
  of planetesimal mass $m$ and rms eccentricity $e_0$. The parameter
  $\mu\equiv \Sigma a^2/M_\odot \sim M_{\rm disc}/M_\odot$ is approximately
  the disc mass relative to the stellar mass.  The region to the right of the
  green, vertical solid line, labelled `$N<1$', is not allowed because the
  planetesimal mass exceeds the assumed disc mass.  The slanted solid line
  divides hot discs, in which planetesimal orbits cross, from cold discs
  (equation [\ref{eq:hot}]).  The regions labelled `unstable'
  represent hot discs that are unstable (in pink) and cold discs that are
  unstable (in blue) according to the approximate equations derived at the end
  of \S\ref{subsect:grav_stab}.  The slanted dashed line separates hot discs
  in which the encounters are shear-dominated from those in which encounters
  are dispersion-dominated.  The pink and blue shaded regions represent the
  allowed parameter values for hot and cold discs, respectively.}
\label{fig:stab}
\end{figure}

The relation between the stability criteria for hot and cold discs can be
clarified using Figure \ref{fig:stab}, which plots allowed regions as a
function of the dimensionless parameters $m/M_\odot$ (ratio of planetesimal
mass to stellar mass; horizontal axis) and $e_0$ (rms eccentricity; vertical
axis). The axes are logarithmic. We aim for clarity at the sake of accuracy by
neglecting all factors of order unity for the rest of this subsection. The
diagram shows the following constraints: 

(i) There must be at least one planetesimal in the disc (equation
[\ref{eq:number}]), so $m/\Sigma a^2 \lesssim 1$ or $m/M_\odot \lesssim \mu$
where $\mu$ is defined in equation (\ref{eq:dimen}). The boundary $m/M_\odot = \mu$ 
is represented by a green, vertical solid line; the excluded region to
the right of this line is labelled `$N<1$'.

(ii) The division between hot and cold discs (\ref{eq:hot}) may be written $e_0 \lesssim m/\mu M_\odot$, which is marked by a slanted solid line.

(iii) The condition (\ref{eq:hillstabb}) for the gravitational stability of
cold discs is $m/M_\odot \gtrsim \mu^{1.43}$, which appears in the figure as a
vertical dotted line; the unstable region to the left of this line is labelled
`unstable' (in blue).

(iv) The condition (\ref{eq:tt}) for gravitational stability of hot discs
becomes $e_0 \gtrsim \mu$ if $m/M_\odot \lesssim \mu^2$.  The second of
equations (\ref{eq:tt}) is neglected because it applies only over a range of a
factor of two in planetesimal mass.  The unstable region is bounded by a solid
horizontal line and labelled `unstable' in pink.  Although this derivation was
carried out for hot discs, it should apply to cold discs as well so long as
$m/M_\odot \lesssim \mu^2$; however, it adds no new restrictions on cold discs
since since these are already unstable by condition (\ref{eq:hillstabb}). 

The pink and blue shaded regions represent the allowed parameter values for
hot and cold discs, respectively. 

\subsubsection{Collisions}
\label{subsect:dynamics_collisions}

The collision time in a system with isotropic velocity dispersion $\sigma$ and
number density $n$ is given by (\citealt{bt08}, equation [7.195])
\be
t_{c}^{-1} = 16 \sqrt{\pi} n \sigma r^2 \left( 1 +  \Theta \right),
\label{eq:tcoll}
\ee
where 
\be
\Theta  \equiv \frac{Gm}{2\sigma^2 r}
\label{eq:safronov}
\ee
is the Safronov number. The factor $(1+\Theta)$ reflects the enhancement in the
collision rate due to gravitational focusing. If the velocity-dispersion
tensor is anisotropic, we define $\Theta$ by replacing $\sigma$ by the radial
dispersion $\sigma_r$, and then with equation (\ref{eq:vdisp}) we have
\be
\Theta={m\over M_\odot}{a\over r}{1\over e_0^2}=1.61 \left({m\over
    M_\odot}\right)^{2/3}{\nu^{1/3}\over e_0^2}
\label{eq:thetaalt}
\ee
where the dimensionless density $\nu$ is defined in equation (\ref{eq:dimen}).

Equation (\ref{eq:tcoll}) requires several corrections for use in discs.  Firstly, the
velocity-dispersion tensor in a Keplerian disc is not isotropic; for the
Rayleigh distribution (\ref{eq:rayleigh}) and our choice $i_0/e_0=0.5$, this
requires replacing $\sigma$ with the radial dispersion $\sigma_r$ and the
factor $(1+\Theta)$ with $(f_1+f_2\Theta)$ where $f_1=0.690$ and $f_2=1.521$.
Secondly, we must account for the falloff in density above the disc midplane,
which we do by replacing the number density $n$ in equation (\ref{eq:tcoll})
with the number-weighted average $\int n^2(z) dz/\int n(z) dz=n_0/\sqrt{2}$,
where $n(z)$ is given by equation (\ref{eq:nz}). Thus, we replace equation (\ref{eq:tcoll}) by
\begin{equation}
\begin{split}
t_{c}^{-1} &= 2^{7/2}\pi^{1/2} n_0 \sigma_r r^2 \left( f_1 + f_2\Theta \right)\\
&= 16{\cal N}\Omega r^2\left(f_1 + f_2\Theta \right)\\
\end{split}
\label{eq:tcolla}
\end{equation}
for $i_0/e_0=0.5$, as we assume throughout.  See Appendix \ref{append:dt93} for details on how to modify the more general formulae of \cite{dt93} to arrive at the preceding result.

Equation (\ref{eq:tcolla}) neglects the Keplerian shear in the disc and thus is only
valid for dispersion-dominated encounters, for which 
\be
\sigma_r \gtrsim \Omega\hbox{\,max}\left\{r, r_{\rm H} \right\},
\label{eq:tcollv}
\ee
where the Hill radius $r_{\rm H}$ is defined in equation (\ref{eq:hillrad}). 
Note that $r_{\rm H}>r$ if the dimensionless parameter $\nu$ (equation [\ref{eq:dimen}])
exceeds 0.7, which is almost always true, so in practice condition
(\ref{eq:tcollv}) reduces to 
\be 
  \sigma_r \gtrsim \Omega r_{\rm H} \quad\hbox{or}\quad e_0\gtrsim (m/M_\odot)^{1/3}.
\label{eq:tcollvv}
\ee
This boundary is marked as a slanted dashed line in Figure \ref{fig:stab}. The allowed
region in parameter space below this line, in which encounters are
shear-dominated, is relatively small but still requires consideration. 
Note that condition (\ref{eq:tcollvv}) can be rewritten with the help of
(\ref{eq:thetaalt}) as $\Theta \lesssim \nu^{1/3}$ so shear-dominated
encounters occur only if the rms eccentricity is so small that the Safronov
number is greater than unity by the large factor $\nu^{1/3}$.  When the condition in equation (\ref{eq:tcollvv}) is violated,
the disc is so flat that encounters may excite out-of-plane motions less efficiently than in-plane
motions, so that $i_0/e_0$ may be less than our assumed value of 0.5. We do
not, however, attempt to model variations in $i_0/e_0$ in this paper. 

Formulae for the shear-dominated collision rate are given by \cite{gl92} and \cite{dt93}.  Adapting these formulae to the present model and notation (see Appendix \ref{append:dt93}), we have
\be
t_{c}^{-1} =
\begin{cases}
f_4 {\cal N} \Omega^2 a^2 r \sigma_r^{-1} \left(m/M_\odot\right)^{2/3}, & \Omega r_{\rm H}/\nu^{1/6} \lesssim \sigma_r \lesssim \Omega r_{\rm H},\\
f_5 {\cal N} \Omega r^{1/2} a^{3/2} \left(m/M_\odot\right)^{1/2}, & \sigma_r \lesssim \Omega r_{\rm H}/\nu^{1/6},\\
\end{cases}
\label{eq:tcollshear}
\ee
where $f_4=22.67$, $f_5=12.94$, and $\nu$ is defined in equation (\ref{eq:dimen}).

Collisions between equal-mass planetesimals may have various outcomes. If the
gravitational binding energy of the planetesimals is negligible compared to
their relative kinetic energy (Safronov number $\Theta\ll1$) then (i)
high-speed collisions will shatter the planetesimals and disperse the
fragments, while (ii) low-speed collisions will leave the planetesimals
unaffected or produce small craters. If the gravitational binding energy is
much larger than the kinetic energy then (iii) high-speed collisions will
still shatter and disperse the planetesimals, while (iv) the outcome of
low-speed collisions will be a single gravitationally bound object containing
most of the mass of the two planetesimals (see \S\ref{sec:coleff} for further
detail). For our purposes, any of the outcomes (i), (iii), or (iv) leads to a
substantial change in the mass distribution of planetesimals if the collision
time is less than the age of the disc, $t_{c}\lesssim t_0$ -- and therefore is
inconsistent with our requirement that the disc is long-lived in its present
state. 

However, many collisions of type (ii) could occur without substantially
altering the planetesimal mass distribution. We therefore differentiate
between dynamically hot discs, in which the collision time is longer than the
age, $t_c\gtrsim t_0$, and `warm' discs, in which the collision time is
shorter than the age but the velocity dispersion $\sigma_r$ is small enough
that the collisions do not substantially damage the planetesimals over the
lifetime of the disc. The survival criteria for warm discs are discussed in
\S\ref{sec:warm}. 

\subsubsection{Gravitational scattering}

The timescale for a substantial change in the rms eccentricity and inclination
due to gravitational scattering of planetesimals must also be longer than the
disc age. We write this as $t_{g} > t_0$, where 
\begin{eqnarray}
t_{g}^{-1}&=&{d\log e_0^2\over dt}\nonumber \\
             &=&{\Omega\over e_0^4}{\mu m\over M_\odot} \left[ S_1(i_0/e_0) ~{\cal C} + S_2(i_0,e_0,m/M_\odot) \right].
\label{eq:tgrav}
\end{eqnarray}

Here the term $S_1(i_0/e_0) {\cal C}$ incorporates the effects of close or
dispersion-dominated encounters, in which the relative velocity is dominated
by the velocity dispersion $\sigma_r$, while $S_2(i_0,e_0,m/M_\odot)$
represents the effects of distant or shear-dominated encounters. The factor
${\cal C}$ is usually called the Coulomb logarithm. We evaluate these terms by
specializing the formulae of \cite{si00} to the case where the colliding
planetesimals have the same mass and the same eccentricity and inclination
distributions. {}From equations (3.29), (3.31) and (6.6)--(6.8) of that paper,
we find $S_1(0.5)=4.50$ and
\begin{eqnarray}
2{\cal C} &=& \log(\Lambda^2+1)-\log(\Lambda_c^2+1)+{1\over \Lambda^2+1}
-{1\over \Lambda^2_c+1}, \nonumber\\
\Lambda&=&{M_\odot\over
  m}(e_0^2+i_0^2)\big[\sqrt{2}i_0+(\ffrac{2}{3}m/M_\odot)^{1/3}\big], \nonumber \\
\Lambda_c&=&{M_\odot\over
  m}{2r\over a}(e_0^2+i_0^2) \left[1+{ma\over M_\odot
    r(e_0^2+i_0^2+\half(\frac{2}{3}m/M_\odot)^{2/3})}\right]^{1/2}\!\!.  
\end{eqnarray}
{}From equations (4.8) and (6.5) of \cite{si00},
\be
S_2(i_0,e_0,m/M_\odot)={3.8\xi \over
  1-i_0^2/e_0^2}[W(\epsilon/e_0^2)-W(\epsilon/i_0^2)], 
\ee
where $\epsilon\equiv \half \xi (\frac{2}{3}m/M_\odot)^{2/3}$,
\be
W\left(Y\right) \equiv \int_Y^\infty \exp(Y-y){dy\over y},
\ee
and $\xi \simeq 2$ is determined by an empirical fit to N-body simulations. 

Using equation (\ref{eq:thetaalt}), equation (\ref{eq:tgrav}) may be rewritten as
\be
t_{g}^{-1}= {\cal N}\Omega r^2\Theta^2 S_1 {\cal C}^\prime,
\label{eq:tgrava}
\ee
where 
\be 
{\cal C}^\prime \equiv {\cal C}+S_2/S_1
\ee
is the correction factor arising from the Coulomb logarithm and
shear-dominated encounters.

The characteristic timescales (\ref{eq:tcolla}) and (\ref{eq:tgrava}) for
physical collisions and gravitational scattering can be combined into a single
relaxation timescale,
\be
t_{\rm relax}^{-1}=t_c^{-1}+t_g^{-1}=16{\cal N}\Omega r^2\left(f_1 + f_2\Theta +
  f_3 {\cal C}^\prime \Theta^2\right),
\label{eq:trelax}
\ee
where $f_3=S_1/16=0.28$. A long-lived disc must have $t_{\rm relax}\gtrsim t_0$.

\begin{figure}
\begin{center}
\includegraphics[width = 0.8 \columnwidth]{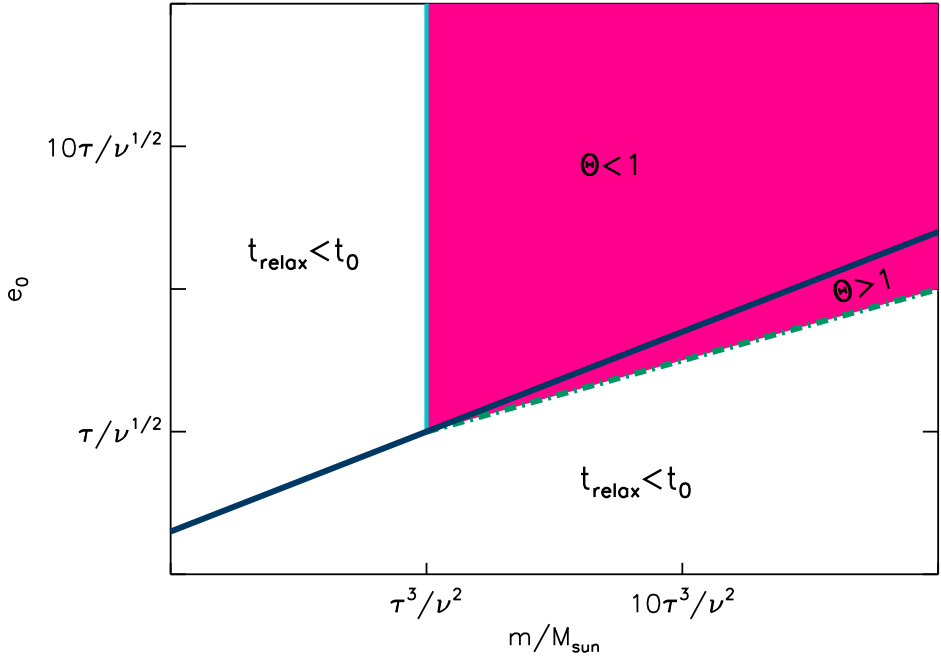}
\end{center}
\caption{Survival of hot discs as a function
  of planetesimal mass $m$ and rms eccentricity $e_0$. The parameters
  $\tau$ and $\nu$ are defined in equation (\ref{eq:dimen}). The slanted solid line divides discs according 
to their Safronov number $\Theta$.  The regions marked by `$t_{\rm relax} < t_0$' are not allowed
  because the collision time (when $\Theta < 1$) or the
  gravitational scattering time (when $\Theta>1$) is shorter than the disc
  age.  The pink shaded region represents the allowed parameter values for hot discs.}
\label{fig:coll}
\end{figure}

The effects of collisions and gravitational scattering can be clarified using
Figure \ref{fig:coll}, which plots allowed regions as a function of
$m/M_\odot$ and $e_0$, in logarithmic coordinates. We neglect all
factors of order unity for the rest of this subsection. In this approximation, the requirement for survival of hot discs simplifies to
\be
(\Omega t_0)^{-1} \gtrsim {\Sigma\over
  m^{1/3}\rho_p^{2/3}}\max\left\{1,\nu^{2/3} \left(\frac{m}{M_\odot}\right)^{4/3} e_0^{-4}\right\}.
\label{eq:trelaxa}
\ee
This result holds only if the collisions are dispersion-dominated, but if they
are shear-dominated the relaxation time becomes shorter than this formula
would predict so the criterion (\ref{eq:trelaxa}) remains necessary (but not
sufficient) for survival of the disc.

The results can be re-written in terms of the dimensionless time $\tau$ defined
in equation (\ref{eq:dimen}).
Discs with Safronov number $\Theta > 1$ have $m/M_\odot > e_0^3/\surd\nu$ and
the survival criterion (\ref{eq:trelaxa}) is $m/M_\odot \gtrsim \tau^3/\nu^2$ ($\Theta \lesssim 1$) or
$m/M_\odot \lesssim e^4_0/\tau$ ($\Theta \gtrsim 1$).  These constraints are shown in 
Figure \ref{fig:coll}.  The excluded regions
are bounded by the cyan, vertical solid ($m/M_\odot=\tau^3/\nu^2$) and green,
slanted dot-dash ($e^4_0 = m\tau/M_\odot$) lines, and are labelled `$t_{\rm
  relax} < t_0$'.  Notice that for a given value 
of the rms eccentricity $e_0$ or velocity dispersion $\sigma_r$, there is only
a finite range of planetesimal masses or radii in which the relaxation time
exceeds the age: the disc cannot survive if the planetesimal mass is either
too small or too large. This behavior arises because for small masses $m$ or
radii $r$ ($\Theta \ll 1$), we have $t_{\rm relax} \propto r \propto m^{1/3}$
at a given surface density and eccentricity, while for large masses ($\Theta
\gg 1$), we have $t_{\rm relax} \propto m^{-1} \propto r^{-3}$.

\subsection{Warm discs}
\label{sec:warm}

As described in \S\ref{subsect:dynamics_collisions}, planetesimals in warm
discs may suffer collisions but the velocity dispersion is low enough that
these collisions do not destroy the planetesimals. Thus a necessary condition
for the survival of warm discs is that the squared velocity dispersion
$\sigma_r^2$ must be less than $Q^\ast_D$, the energy per unit mass
required to disperse the planetesimal into fragments, which we obtain from
equation (\ref{eq:qast}) below. 

This criterion is not sufficient, for two reasons. Firstly, if the gravitational
binding energy of the planetesimals is much larger than the kinetic energy,
an inelastic collision is likely to leave the colliding planetesimals as a
gravitationally bound pair, which alters the mass distribution and therefore
is inconsistent with our assumption that the disc has not evolved. Therefore
we require that the Safronov number $\Theta \lesssim 1$ for warm discs. 

Secondly, low-velocity collisions can chip or crater the planetesimals even if
they are not disrupted. Thus if the collision time $t_c$ is much less than the
age $t_0$ the planetesimals may be gradually eroded away even if they are not
dispersed in a single collision. A simple parametrization of the erosive
process is to assume that the mass lost in a typical collision of two
objects of mass $m$ at relative velocity $v$ is \citep[e.g.,][]{ta07}
\be
\Delta m=0.5 m\left( v^2\over Q_D^\ast\right)^\gamma, \quad v^2 \lesssim Q_D^\ast
\ee
(the factor 0.5 arises because the usual definition of a dispersive impact is
one in which the mass of the largest fragment is less than half of the target
mass; see \S\ref{sec:coleff}). 

Erosion through impact involves a number of complicated processes such as
crack propagation in brittle materials and plastic flow in ductile materials,
and melting or sublimation at high impact velocities. Our present state of
knowledge is derived from a variety of approximate physical models,
experiments, and numerical simulations, most intended for situations far
removed from planetesimal collisions. These typically yield values of
$\gamma$ between 1 and 1.5. For example, (i) impacts of small pellets into
targets composed of ice-silicate mixtures at speeds up to $12\kms$ yield
$\gamma\simeq 1.2$ \citep{kg01}; (ii) numerical simulations of collisions between
rocky and icy bodies yield $\gamma\simeq 1$ \citep{ba99,sl09}. We shall adopt
$\gamma=1$ recognizing that this is (a) oversimplified; (b) conservative (in
that the actual erosion rate is expected to be smaller for low-velocity
collisions if $\gamma$ is larger). 

The criterion for survival is that the cumulative mass loss $\Delta m(t_0/t_c)
\lesssim m$. Thus warm discs must satisfy
\be
t_c\lesssim t_0,\quad \Theta\lesssim 1, \quad \mbox{and}\quad \sigma_r^2 \lesssim
\left(t_c/t_0 \right) Q^\ast_D. 
\label{eq:hotdiscs_type}
\ee 
where $Q^\ast_D$ is given by equation (\ref{eq:qast})\footnote{The distinction
  between `hot' and `warm' discs based on whether or not the collisions
  are destructive is moot when the collision time is longer than the age. Our
  (arbitrary) convention is that such discs are `hot', not `warm'.}.

An additional constraint for warm discs is that collisions do not result in
excessive `viscous' spreading of the disc.  A disc initially localized at
semi-major axis $a$ that loses a small amount of energy $\delta E$ must spread
by $\delta a \ll a$ \citep{bra77}, where 
\be
\delta E = -\frac{G M_\odot M_{\rm disc}}{32 a^3} \delta a^2.
\ee
If the typical energy lost per unit mass in a collision is $f_6 \sigma^2_r$,
where $f_6 \sim 1$, then the rate of energy loss is $f_6 \sigma^2_r M_{\rm
  disc}/t_c$.  Therefore, in a time $t_0$ the disc spreads to 
\be
\frac{\delta a}{a} = \sqrt{\frac{32 f_6 a \sigma^2_r}{G M_\odot}\frac{t_0}{t_c}}.
\label{eq:spread}
\ee
Requiring $\delta a/a \lesssim 1$ and assuming $\Theta \ll 1$ yields
\be
\sigma_r \lesssim 8 \times 10^2 \mbox{ cm s}^{-1}
\,\ \left(\frac{f_m}{f_6}\right)^{1/2} \left(\frac{f_1}{0.69}\right)^{-1/2} \left(\frac{\rho_p}{3\gmcm}
  \frac{r}{10^5 \mbox{ cm}}\right)^{1/2} \left( \frac{\mu}{10^{-4}}
  \frac{t_0}{3\Gyr} \right)^{-1/2} \left(\frac{a}{10\au}\right)^{5/4}.
\label{eq:viscous}
\ee
This constraint does not restrict the allowed range for warm discs among
the six sample discs considered in this paper, i.e., it is satisfied so long
as the warm discs satisfy all of the other constraints we have already discussed.

\section{Collision-limited discs}
\label{sec:col}

\subsection{The effects of collisions}
\label{sec:coleff}

Collisions or impacts may crater, shatter or disperse the target, as discussed
in \S\ref{subsect:dynamics_collisions} and \S\ref{sec:warm}.  The distinction
between shattering and dispersive impacts arises because the pieces of the
target may remain gravitationally bound even after the target is shattered
\citep{ba99}.

The outcome of an impact depends mainly on the relative speed, masses, and
composition of the impacting bodies. Consider an impact between two
planetesimals of masses $m_1$ (the target) and $m_2\le m_1$ (the projectile),
at relative velocity $\Delta v$. The kinetic energy of relative motion is
$E_k=\half \tilde{m} \Delta v^2$ where $\tilde{m} \equiv m_1m_2/(m_1+m_2)$ is
the reduced mass. The impact is shattering if the kinetic energy per unit
target mass $E_k/m_1>Q^\ast_S$, and dispersive if $E_k/m_1>Q^\ast_D$, where
$Q^\ast$ is a function of the mass and composition of the target. Roughly
speaking, $Q^\ast(m)$ is the binding energy per unit mass of the target --
the energy needed to rupture its internal chemical bonds for $Q^\ast_S$ and
this energy plus its gravitational potential energy for $Q^\ast_D$. In the
strength-dominated regime, when the target is small enough that its
self-gravity is negligible, we expect $Q^\ast_S\sim Q^\ast_D$, and both should
be independent of $m$ in the idealized case in which the strength of the
target is independent of its size. In practice, both simulations and
experiments find that $Q^\ast$ declines slowly with $m$ in the
strength-dominated regime. In the gravity-dominated regime, $Q^\ast_S\ll
Q^\ast_D$ (an impact that shatters a target may not impart escape speed to the
fragments, so they reaccumulate as a gravitationally bound rubble pile) and we
might expect that $Q^\ast_D\propto m^{2/3}$ since the gravitational binding
energy per unit mass of a homogeneous body scales as $m^{2/3}$.

The dependence of $Q^\ast_D$ on target mass is typically parametrized as
\be
Q^\ast_D(m)=Q_0\left(x^a_0+ x^b_0\right),
\label{eq:qast} 
\ee
where $x_0 \equiv m/m_0$; $Q_0$, $m_0$, $a$ and $b$ are parameters to be fitted to experiments or
simulations. We shall adopt $m_0=10^{14}\gm$, $Q_0=6\times
10^5\,\hbox{erg g}^{-1}$, $a=-0.13$ and $b=0.44$, these values being compromises
between the results for ice and basalt given by \cite{ba99}. This simple form
fails for very small mass, since it predicts $Q^\ast_D(m)\to\infty$ as $m\to
0$; experiments with high-velocity impacts of small bodies suggest
$Q^\ast_D\approx 10^7\,\hbox{erg g}^{-1}$ independent of mass \citep{fly04},
so for $x_0 \ll 1$ we use the smaller of this value and the prediction of equation
(\ref{eq:qast}) -- the transition occurs at $m \approx 10^4\gm$.

\subsection{The collisional cascade} 
\label{sec:cc}

When the collision time is much less than the age of the disc, the mass
distribution of the smaller planetesimals is likely to be established by a
`collisional cascade', in which large bodies are dispersed by collisions
into smaller bodies, these in turn being dispersed into smaller ones, until
bodies of size $\lesssim 1\,\mu$m (`dust grains') are removed by
Poynting--Robertson drag and radiation pressure (see \S\ref{sec:dust}). We now
derive an approximate form for the steady-state mass distribution in a
collisional cascade \citep{doh69,og03,ps05}, using the following assumptions:

\begin{enumerate}

\item The number density of planetesimals per unit mass is a power-law function of mass, at
  least over a limited range, 
\be
 \frac{dn}{dm}   \propto m^{-p}.
\label{eq:massdist}
\ee

\item The rms eccentricity of the planetesimals is independent of mass.  (This
  assumption is made for simplicity, but appears to hold approximately in
  the asteroid and Kuiper belts.)

\item The velocity dispersion (\ref{eq:vdisp}) is large enough that a typical
  impact between two bodies of equal mass has much more than enough energy to
  disperse the two bodies, i.e., $\sigma_r^2 \gg Q^\ast_D$, which requires
  $\sigma_r\gg 3 \times 10^3\,\hbox{cm 
    s}^{-1}(Q^\ast_D/10^7\,\hbox{erg g}^{-1})^{1/2}$ or $e_0\gg
  0.005(a/10\au)^{1/2}\*(Q^\ast_D/10^7\,\hbox{erg g}^{-1})^{1/2}$.

\item The specific impact energy required for a dispersive impact is a
  power-law function of mass, $Q^\ast_D \propto m^j$. According to equation
  (\ref{eq:qast}), we expect $j=a$ for $m\ll m_0$ (strength-dominated) and
  $j=b$ for $m\gg m_0$ (gravity-dominated).

\item The cross-section for a collision between two bodies of masses $m_1$ and
  $m_2\ll m_1$ is $\sigma_{\rm coll}(m_1,m_2) \propto m_1^l$. {}From equation
  (\ref{eq:tcoll}) we expect $l=\frac{2}{3}$ for $\Theta\ll 1$ ($\sigma_{\rm
    coll} \propto r^2\propto m^{2/3}$) and $l=\frac{4}{3}$ for $\Theta \gg 1$
  ($\sigma_{\rm coll} \propto r^2\Theta \propto rm \propto m^{4/3}$).

\end{enumerate}

According to assumption (iii), the smallest projectile mass that can disperse a
target of mass 
$m$ in an impact is $m_{\rm min}(m)\approx m Q^\ast_D/\sigma_r^2 \ll m$. The rate
of collisions in which planetesimals of mass $>m$ are dispersed is approximately
\be
\Phi_n(m) \propto \int_m^\infty dm_1 \int_{m_{\rm min}(m_1)}^{m_1} dm_2
~\frac{dn\left(m_1\right)}{dm_1} \frac{dn(m_2)}{dm_2} \sigma_{\rm
  coll}\left(m_1,m_2\right).
\ee
This formula is accurate to a factor of order unity only, since it neglects
the fact that one object of mass $>m$ is dispersed when $m_1>m$ and $m_2<m$,
while two are dispersed when $m_1,m_2>m$; it also neglects the possibility
that one or more of the collision fragments is more massive than $m$. These
inaccuracies do not affect our final result. 
 
Since most of the fragments in the collisions that dominate this rate will
have masses $< m$, the mass per unit volume in planetesimals larger than $m$
decreases at a rate given approximately by 
\be
\Phi_m(m) \approx  m\Phi_n(m) \propto  m\int_m^\infty dm_1 \int_{m_{\rm min}(m_1)}^{m_1} dm_2 \frac{dn\left(m_1\right)}{dm_1} \frac{dn\left(m_2\right)}{dm_2} \sigma_{\rm coll}\left(m_1,m_2\right).
\ee
With assumptions (i), (iv), and (v), we find
\be
\Phi_m(m) \propto m^{3+j+l-p(2+j)}.
\ee
In a steady state, the mass flux $\Phi_m(m)$ must be independent of mass, so
\be
p={3+l+j\over 2+j}.
\ee
With the parameters used in this paper, almost all dispersive collisions have
$\Theta \ll 1$ so we can set $l=\frac{2}{3}$ and obtain \citep{og03}
\be
p={11+3j\over 6+3j}.
\label{eq:massexp}
\ee
Matching the power-law behavior for $m\ll m_0$ and $m\gg m_0$, we have \citep{loh08}
\be
\frac{dn(m)}{dm}=\left\{\begin{array}{ll} (m/m_0)^{-(11+3b)/(6+3b)}, & \mbox{ $m>m_0$}, \\
                              (m/m_0)^{-(11+3a)/(6+3a)}, & \mbox{ $m<m_0$}.
    \end{array}\right.
\ee
These expressions are only valid if (i) the collision time at mass $m$ is
short compared to the age of the disc; (ii) the typical impact velocity is
sufficient to disperse a body of mass $m$, $\sigma_r^2\gg Q^\ast_D(m)$; (iii)
the velocity dispersion is independent of mass; (iv) 
the mass $m$ is sufficiently large that the lifetime to radiation pressure and
Poynting-Robertson drag is much larger than the collisional lifetime.

For the exponents $a=-0.13$, $b=0.44$ given after equation (\ref{eq:qast}), we
have $p=1.89$ for $m\ll m_0$ and $p=1.68$ for $m\gg m_0$. With these exponents the
total mass in the collisional cascade ($\propto \int m \,dn \propto
m^{2-p}$) is dominated by the largest bodies, while the total cross-section
($\propto \int m^{2/3} \,dn \propto m^{5/3-p}$) is dominated by the smallest
bodies. 

\subsection{Properties of collision-limited discs}
\label{subsect:coll_limited}

Collision-limited discs can arise if we assume that there is an initial
distribution of planetesimal masses in which the total disc mass is dominated
by small bodies, say $dn \propto m^{-P}dm$ with $P>2$.\footnote{The condition
  $P\gtrsim 2$ appears to hold for most planetesimal populations in the
  solar system. For the classical and excited Kuiper belts, $P\simeq 3.3$ and
  $2.1$ respectively, and for trans-Neptunian objects $P\simeq 2.5$
  \citep{bern04}. Numerical models of the formation of the Kuiper belt give
  $P=2.3$ \citep{kb04}.  Asteroid observations yield $P$ between 1.9 and 2.3
  \citep{par08} and Jupiter-family comets have $P=1.9$ though with large
  uncertainties \citep{fer05}.} In such discs the collision time is shorter
for planetesimals of smaller mass; thus a collisional cascade is established
for all masses below some maximum $m_{\rm max}$. We have seen that the mass in
the collisional cascade is dominated by the largest bodies in the cascade,
while if $P>2$ the mass in the `primordial' regime $m>m_{\rm max}$ is
dominated by the smallest bodies. Thus, the overall mass in the disc is
dominated by bodies with mass $\sim m_{\rm max}$, and for many purposes we may
treat the disc as a monodisperse system composed of bodies with a single mass
$m_{\rm max}$. The difference from our earlier discussions is that now the
planetesimal mass is not a free parameter; rather, it is determined by the
condition that the lifetime of a planetesimal of mass $m_{\rm max}$ subject to
dispersive impacts is equal to the disc age. We now derive this condition.

Following the discussion in the preceding section, we shall assume that
dispersive impacts have Safronov number $\Theta \ll 1$ and are dominated by
impactors of mass much less than the target mass. Then if we equate the
collision time from equation (\ref{eq:tcolla}) to the disc age $t_0$,
we have 
\be
t_0^{-1}=8\sqrt{2\pi}f_1\sigma_r \left(3m_{\rm max}\over
  4\pi\rho_p\right)^{2/3}\! \int_{m_{\rm min}(m_{\rm max})}^{m_{\rm max}}\!\!\! \frac{dn_0\left(m'\right)}{dm'}dm',
\label{eq:tcollc}
\ee
where $m_{\rm min}=m_{\rm max} Q^\ast_D(m_{\rm  max})/\sigma_r^2$ is the smallest impactor that will disperse a target of
mass $m_{\rm max}$. Assuming that the mass distribution in the midplane of the
disc is 
\be
\frac{dn_0\left(m\right)}{dm}=\frac{n_{\rm max}}{m_{\rm max}}
\begin{cases}
\left(m/m_{\rm max}\right)^{-p} &  \quad m \le m_{\rm max},\\
\left(m/m_{\rm max}\right)^{-P} &  \quad m \ge m_{\rm max},\\
\end{cases}
\ee
we have
\be
t_0^{-1} \simeq {5.32\over p-1}\frac{f_1}{0.69} {n_{\rm max} m_{\rm max}^{2/3}\sigma_r\over
  \rho_p^{2/3}}\left[Q^\ast_D(m_{\rm max})\over\sigma_r^2\right]^{1-p}.
\label{eq:tcolle}
\ee
The total midplane mass density $\rho_0$ is related to the surface density by
$\Sigma=\sqrt{\pi}\rho_0 ai_0$ (equation [\ref{eq:rhosig}]) and $\rho_0$ is given
by 
\begin{eqnarray}
\rho_0&=& \frac{n_{\rm max}}{m_{\rm max}} \left( m_{\rm max}^p \int_0^{m_{\rm max}}\!\!m^{1-p}dm +
   m_{\rm max}^P \int_{m_{\rm max}}^\infty\!\! m^{1-P}dm\right) \nonumber \\
    &=& n_{\rm max} m_{\rm max} \left( \frac{1}{2-p} + \frac{1}{P-2} \right),
\label{eq:tcolld}
\end{eqnarray}
assuming $p<2<P$. If we divide the second equation
by the first and use equations (\ref{eq:rhosig}) and (\ref{eq:vdisp}) we obtain
\be
\Omega t_0 \simeq 1.25 f_7 \left(\frac{f_1}{0.69} \right)^{-1} ~{m_{\rm max}^{1/3}\rho_p^{2/3}\over\Sigma}
\left[Q^\ast_D\left(m_{\rm max}\right)\over \sigma_r^2\right]^{p-1},
\label{eq:abc}
\ee
which is an implicit equation for the characteristic planetesimal mass $m_{\rm
max}$. Here $f_7\equiv (p-1)[1/(2-p)+1/(P-2)]/5.32$. We choose $p=1.68$ and
$P=4$ so $f_7=0.46$; a steeper high-mass slope $P=6$ would change this only to
$f_7=0.43$. Alternatively, we may write
\be
\sigma_r \simeq \left[ 1.25 f_7 \left(\frac{f_1}{0.69} \right)^{-1} \frac{m_{\rm
      max}^{1/3}\rho_p^{2/3}}{\Sigma \Omega t_0} \right]^{\frac{1}
      {2\left(p-1\right)}} \sqrt{Q^\ast_D\left(m_{\rm max}\right)}. 
\label{eq:vcoll_limited}
\ee
By replacing the mass $m$ in our discussion of monodisperse discs with $m_{\rm
  max}$, most of the results of \S\ref{sec:simple} and \S\ref{sec:mono} can be
applied to collision-limited discs without major errors. For example: (i) The
gravitational stability of hot discs depends on the surface density $\Sigma
\propto \int m \,dn$, which is dominated by masses near $m_{\rm max}$ when
$p<2<P$. (ii) For most purposes the appropriate replacement for the collision
rate $t_c^{-1}\propto nr^2(f_1+f_2\Theta)$ (equation (\ref{eq:tcolla}]) in a
disc with a distribution of masses is the mass-weighted collision rate
$\propto \int m r^2(f_1+f_2\Theta) \,dn$, which in turn is proportional to
$\int m^{5/3} \,dn$ for $\Theta\ll1$ and $\int m^{7/3} \,dn$ for
$\Theta\gg1$. These integrals are dominated by masses near $m_{\rm max}$ when
$p<8/3<P$ and $p<10/3<P$ respectively. (iii) The gravitational scattering rate
is $t_g^{-1}\propto \int r^2\Theta^2 \,dn \propto \int m^2 \,dn$ so
this is dominated by masses near $m_{\rm max}$ when $p<3<P$. All of these
inequalities are satisfied for our nominal values $p=1.68$ and $P=4$.

Finally we note an interesting inconsistency in the results we have derived so
far. Neglecting factors of order unity, the collision time $t_{\rm c,m}$ for a
monodisperse disc with $\Theta\ll1$ is given by equation (\ref{eq:tcolla}) as
$t_{\rm c,m}^{-1}\sim n_0\sigma_rm^{2/3}/\rho_p^{2/3}$. The analogous collision
time $t_{\rm c,cl}$ for a collision-limited disc is given by equation
(\ref{eq:tcolle}) as $t_{\rm c,cl}^{-1}\sim n_{\rm max}\sigma_rm_{\rm
  max}^{2/3}/\rho_p^{2/3}(Q_D^\ast/\sigma_r^2)^{1-p}\sim t_{\rm c,m}^{-1}(Q_D^\ast/\sigma_r^2)^{1-p} \gtrsim t_{\rm c,m}^{-1}$. The difference arises
because the collision-limited disc has a large population of bodies with
$m_{\rm min}\lesssim m \lesssim m_{\rm max}$ that can collide with and
disperse the large planetesimals, whereas in the monodisperse disc these are
destroyed only by collisions among themselves. What then is the state of a
disc in which $t_{\rm c,cl}\lesssim t_0 \lesssim t_{\rm c,m}$? Should it be
regarded as monodisperse or collision-limited? We assume here that such discs
are monodisperse but this assumption may be oversimplified. 

\section{Non-gravitational forces on dust}
\label{sec:dust}

We also describe the most important non-gravitational forces on
dust grains \citep{bu79}, since the distribution of dust grains determines the
infrared flux from debris discs; these forces can also be relevant for the
planetesimals in warm discs.  
Gas drag on the dust is unimportant since we
are focusing on discs older than a few Myr, at which point the gas in the
protoplanetary disc has disappeared. The ratio of repulsive forces from the
stellar wind and radiation pressure to the attractive gravitational force is
\cite[e.g.,][]{sc06}
\be
\beta ={3\over 16\pi}{L_\odot {\cal P}_r\over GM_\odot
  c\rho_pr}=0.19\,{\cal P}_r \left(\frac{\rho_p}{3\gmcm}\right)^{-1} \left(\frac{r}{1\,\mu\mbox{m}}\right)^{-1},
\label{eq:beta}
\ee
in which we have assumed that the host star has the solar mass $M_\odot$ and
luminosity $L_\odot$, $\rho_p$ is the dust grain density, $r$ is the grain
radius, and 
\be
{\cal P}_r=Q_{pr}+{\dot Mv_wc\over L_\odot},
\ee
where $\dot M$ is the rate of mass loss in the stellar wind, $v_w$ is the wind
speed, and $Q_{pr}$ is the radiation pressure efficiency factor (averaged over
the stellar spectrum) as defined by \cite{bu79}.  For stars with the Sun's
luminosity and age, the contribution of the stellar wind to ${\cal P}_r$ is
negligible ($\lesssim 10^{-3}$). Grains created by collisions on circular
orbits with the Keplerian speed are unbound if $\beta\ge \half$ -- because
their eccentricity is $\beta/(1-\beta)$ \citep{bu79} -- and thus $\beta=\half$
defines the `blow-out radius',
\be
r_b=0.38\,\mu\mbox{m} ~{\cal P}_r\left(\frac{\rho_p}{3\gmcm}\right)^{-1}.
\label{eq:r_blowout}
\ee

The rate of orbital decay from Poynting--Robertson and stellar-wind drag is
\citep{bu79} 
\be
{1\over t_{\rm PR}} \equiv -{1\over a}{da\over dt}={3\over 8\pi}{L_\odot {\cal
    P}_\phi\over c^2a^2\rho_pr}={2{\cal P}_\phi\over {\cal P}_r}{GM_\odot\over
  ca^2}\beta, 
\ee
where
\be
{\cal P}_\phi=Q_{pr} + {\dot Mc^2\over L_\odot}.
\ee
Numerically, we have
\be
t_{\rm PR}=8.0\times10^4\hbox{\,yr} ~\beta^{-1} {{\cal P}_r\over {\cal P}_\phi}\left(a\over 10\au\right)^2.
\label{eq:tpr}
\ee
The contribution of the stellar wind to ${\cal P}_\phi$ is negligible for particles
with $r \gtrsim 1\,\mu$m but grows as $r^{-1}$ for $r\lesssim 0.3\,\mu$m, and equals
the contribution due to Poynting-Robertson drag at $r \approx 0.1\,\mu$m
\citep{bu79}. 

The thermal emission from the dust is determined by its absorption efficiency
factor $Q_a$, which is similar in magnitude to $Q_{pr}$ and equal to it if
scattering is neglected. In the geometric optics limit, $Q_a$ is independent
of wavelength and close to unity for typical dust grains; in the limit of long
wavelength, where $X\equiv 2\pi r/\lambda\ll1$, $Q_a\sim X$ (equation
[\ref{eq:mie}]). The contribution of a grain to the thermal emissivity at
wavelength $\lambda$ is proportional to $\pi r^2Q_a$. Thus for a power-law
mass distribution with exponent $-p$ (equation [\ref{eq:massdist}]) in the range
$\frac{5}{3} < p < 2$ (as is the case for a collisional cascade; see
\S\ref{sec:cc})\footnote{The assumption of a power-law mass distribution
  neglects the oscillations that appear in the mass distribution of a
  collisional cascade at radii that are not too far from the blow-out radius
  \citep{kriv06}; these oscillations can change the number density at a given
  radius or mass by a factor of $\sim 3$.} the total thermal emissivity is dominated by grains with
$X=2\pi r/\lambda\sim1$ or $r\simeq r_{\rm IR}(\lambda)\equiv \lambda/(2\pi)$.
More precisely, a fraction $f$ of the emission comes from particles with radii
that exceed $r_{\rm IR}/\kappa$, where $f=1-(3p-5)\kappa^{3(p-2)}$. For
$p=11/6$, the expected value for a collisional cascade (we use $p=11/6=1.83$
rather than $p=1.89$ as at the end of \S\ref{sec:cc} for reasons given after 
equation [\ref{eq:fire}]), 75\% of the emission comes from particles with
radii that exceed $0.25r_{\rm IR}$. For observations at $\lambda=20\,\mu$m,
$r_{\rm IR}=3\,\mu$m and 75\% of the emission comes from particles with radii
that exceed $0.8\,\mu$m, where $\beta=0.24$ (assuming $\rho_p=3\gmcm$ and
$Q_a=1$, since most of the stellar emission is at shorter wavelengths). At
longer observational wavelengths $\beta$ is even smaller for the particles
dominating the emission. Thus radiation pressure is negligible, except perhaps
for accurate modeling at the shortest observational wavelengths.

Poynting--Robertson drag
is also negligible, at least for detectable debris discs, as shown by the
following argument. Assume for simplicity that the dust particles have a
single size. Using equations (\ref{eq:taupdef}) and (\ref{eq:tcolla}) the
collision time $t_c$ for dust is related to the geometrical optical depth as seen
from the host star $\tau_p$ by
\be
\Omega t_c={\pi f_m\over 64 f_1 \tau_p} \quad\mbox{or}\quad t_c=3.6\times
10^4\yr\,f_m\left(f_1\over0.69\right)^{-1} \left({\tau_p\over
    10^{-5}}\right)^{-1} \left(a\over 10\au\right)^{3/2}.
\label{eq:tautcoll}
\ee
For particles with absorption efficiency $Q_a\simeq 1$, the optical depth
$\tau_p$ is equal to the bolometric luminosity of the disc relative to the
star, which exceeds $10^{-5}$ in almost all observed debris discs
\citep{wy08}. Therefore the collision time $t_c\lesssim 4\times
10^4\yr\,(a/10\au)^{3/2}$ in observed discs, shorter than the
Poynting--Robertson drag time (\ref{eq:tpr}), so the grains are destroyed by
collisions before they experience significant orbital decay. 

Based on these arguments, we neglect non-gravitational forces on the dust
distribution when calculating its emission properties.

\section{The properties of long-lived discs}
\label{sect:longliveddiscs}

The primary goal of this paper is to explore the properties of planetesimal
discs that can survive for most of the age of the Galaxy. Even in our highly
simplified model, the local properties of discs are specified by four
parameters: semi-major axis $a$, surface density
$\Sigma$, planetesimal radius $r$ or mass $m$, and rms eccentricity
$e_0$, while collision-limited discs are specified by three parameters. It is
challenging to visualize the properties of a four-dimensional parameter
space. As a first step, we outline some general conclusions that arise from
the discussion of the previous section.

\subsection{Cold discs}

\subsubsection{The minimum planetesimal mass in cold discs}

Eliminating the separation $\Delta a$ between equations (\ref{eq:da}) and
(\ref{eq:hillstabb}) yields the minimum mass in cold discs,
\be
m > m_{\rm min}^{\rm cold}\equiv 100M_\odot \mu^{1.43} \simeq 60M_\oplus\left( \mu \over 10^{-4} \right)^{1.43}\!\!.
\label{eq:mmincold}
\ee
Thus, the minimum planetesimal mass in a cold disc with $\mu=10^{-4}$ is 0.2 Jupiter masses, while if  $\mu=10^{-6}$ the minimum mass is 0.1 Earth masses. 

\subsubsection{The maximum number of planetesimals in cold discs}

If we write the disc mass as $M_{\rm disc}=\pi f_m \Sigma a^2$ 
(with $f_m=1$ this is roughly the mass per octave in semi-major axis; a
disc extending over multiple octaves in semi-major axis could have $f_m\gg1$),
then with equation (\ref{eq:mmincold}) we have 
\be
N = {M_{\rm disc}\over m} < N_{\rm max}^{\rm cold}\equiv 
{f_m\over 8}\left(M_\odot\over m\right)^{0.3},
\label{eq:nmax}
\ee
or $N\sim 1$ for Jupiter-mass planets, 6 for Earth-mass planets, and 20 for
lunar-mass planets. 

This result can be re-cast in terms of the surface density,\invisible{actual number
  1.626} 
\be
N < N_{\rm max}^{\rm cold}\equiv 1.6f_m\left(\mu \over 10^{-4}\right)^{-0.43}.
\label{eq:nmaxa}
\ee
Thus cold discs with $\mu=10^{-4}$ can host no more than one or two
equal-mass planetesimals per octave in radius, while cold discs with  $\mu=10^{-6}$ can host up to 12.

\subsection{Hot discs}

\label{sec:hotdisc}

\subsubsection{The maximum surface density for hot discs}

Comparing Figures \ref{fig:stab} and \ref{fig:coll} shows that a necessary
condition for the survival of a hot disc is that the allowed regions in the
two figures overlap, and that this requires (i) $\tau^3/\nu^2\lesssim \mu$
(the minimum planetesimal mass for which the collision time is longer than the
age must be smaller than the disc mass) and (ii) $\tau/\nu^{1/2}\lesssim 1$
(the minimum eccentricity for which the collision and gravitational scattering
times are less than the age must be less than unity). The first of these can
be written more accurately using equations (\ref{eq:number}) and
(\ref{eq:tcolla}) as\invisible{actual number 7.020} 
\be
\mu < \mu_{\rm max,c}
\label{eq:exista}
\ee
where
\be
\begin{split}
\mu_{\rm max,c} &=  {\pi^{3/2}f_m^{1/2}\over 48f_1^{3/2}} \frac{\nu}{(\Omega
  t_0)^{3/2}}\\ 
&= 7.0\times 10^{-5} f^{1/2}_m ~\left(\frac{\rho_p}{3\gmcm} \right) \left(\frac{t_0}{3\Gyr} \right)^{-3/2} \left(\frac{a}{10\au} \right)^{21/4}.\\
\end{split}
\ee

To describe condition (ii) more accurately, we re-write the relaxation time
(\ref{eq:trelax}) using equation (\ref{eq:thetaalt}):
\be
t^{-1}_{\rm relax} = {16\Sigma\Omega\over e_0}\left(3a\over 4\pi\rho_pM_\odot\right)^{1/2} (f_1\Theta^{-1/2}+f_2\Theta^{1/2}+f_3\Theta^{3/2}\log\Lambda),
\ee
in which we have assumed $\Lambda\gg 1$, as is usually the case, and neglected
the contribution of shear-dominated encounters. The minimum of the expression
in brackets occurs when $\Theta=f_1^{1/2}/(3f_3\log\Lambda)^{1/2}$ for
$\log\Lambda\gg1$, and equals $0.97(\log\Lambda)^{1/4}$. Since 
$e_0<f_e$ (equation [\ref{eq:thin}]), the relaxation time cannot be greater than 
the age $t_0$ for any rms eccentricity unless
\be
\mu < \mu_{\rm max,relax}
\label{eq:existb}
\ee
where
\be
\mu_{\rm max,relax} = 4.4 \times 10^{-6} ~\left(\frac{f_e}{0.5} \right) \left(\frac{f_1}{0.69}\right)^{-3/4} \left(\frac{f_3}{0.28}\right)^{-1/4} \left(\frac{\log\Lambda}{10} \right)^{-1/4} \left(\frac{\rho_p}{3\gmcm} \right)^{1/2} \left(\frac{t_0}{3\Gyr} \right)^{-1} \left(\frac{a}{10\au} \right)^3.
\ee

The existence of a maximum surface density implies a maximum value for the IR
excess emission due to dust (see \S\ref{subsect:fmax}). 

\subsubsection{The maximum number of planetesimals in hot discs}

When a hot disc satisfies the constraints (\ref{eq:exista}) and
(\ref{eq:existb}), for a given surface density and semi-major axis there is a a
minimum planetesimal mass and maximum number of planetesimals, 
given approximately by (cf.\ Fig.\ \ref{fig:coll})
\be
{m_{\rm min}^{\rm hot}\over M_\odot} \sim {\tau^3\over\nu^2}={(\Sigma \Omega
  t_0)^3\over M_\odot\rho_p^2},\quad N_{\rm max}^{\rm hot} \sim \left(\rho_pa\over
  \Sigma\right)^2{1\over(\Omega t_0)^3}.
\ee
More precisely, by evaluating $t_c = t_0$ ($\Theta \ll 1$) we get\invisible{actual number 0.4927}:
\begin{eqnarray}
m_{\rm min}^{\rm hot}&=& {2304f_1^3\over\pi^2}
  {(\Sigma \Omega t_0)^3\over\rho_p^2} \nonumber \\
  &=& 210 M_\oplus \left(\mu \over 10^{-4} \right)^3 \left( f_1\over 0.69\right)^3
  \left(\rho_p \over 3\gmcm\right)^{-2} \left(t_0\over 3\Gyr\right)^3\left(a \over 10\au\right)^{-21/2},\nonumber \\
N_{\rm max}^{\rm hot} &=& {\pi^3f_m\over 2304 f_1^3}\left(\rho_pa\over\Sigma\right)^2
  {1\over(\Omega t_0)^3} \nonumber \\
  &=& 0.49 f_m \left(f_1\over 0.69\right)^{-3}\left(\mu \over 10^{-4}\right)^{-2} \left(\rho_p\over3\gmcm\right)^2 \left(t_0 \over 3\Gyr\right)^{-3} \left(a\over10\au\right)^{21/2}.
\label{eq:hotdiscc}
\end{eqnarray}

\subsection{Warm discs}

Similarly, for given values of the surface density and semi-major axes there
is a maximum mass ($m_{\rm max}^{\rm warm}$) -- and thereby a minimum number
($N_{\rm min}^{\rm warm}$) -- of planetesimals in warm discs. Generally, the
maximum mass and mininum number of planetesimals need to be evaluated
numerically, but for the cases considered below (\S\ref{sec:sample}), these
extremes are attained when the line $t_c=t_0$ intersects the line $\Theta=1$
(cf.\ Figs.\ \ref{fig:sample_discs2} and \ref{fig:sample_discs3}).
At this point, 
\be
\begin{split}
&m_{\rm max}^{\rm warm} = 0.007 M_\oplus\, \left(\frac{f_1+f_2}{2.211} \right)^3 \left(\mu \over 10^{-6} \right)^3
\left(\rho_p \over 3\gmcm\right)^{-2} \left(t_0\over 3\Gyr\right)^3\left(a
  \over 10\au\right)^{-21/2},\\ 
&N_{\rm min}^{\rm warm} = 150 f_m ~\left(\frac{f_1+f_2}{2.211} \right)^{-3} \left(\mu \over 10^{-6}
\right)^{-2} \left(\rho_p \over 3\gmcm\right)^2 \left(t_0\over
  3\Gyr\right)^{-3} \left(a \over 10\au\right)^{21/2}.\\ 
\end{split}
\label{eq:warmdisc_mmax}
\ee

\subsection{Sample discs}
\label{sec:sample}

\begin{figure}
\begin{center}
\includegraphics[width=0.8\columnwidth]{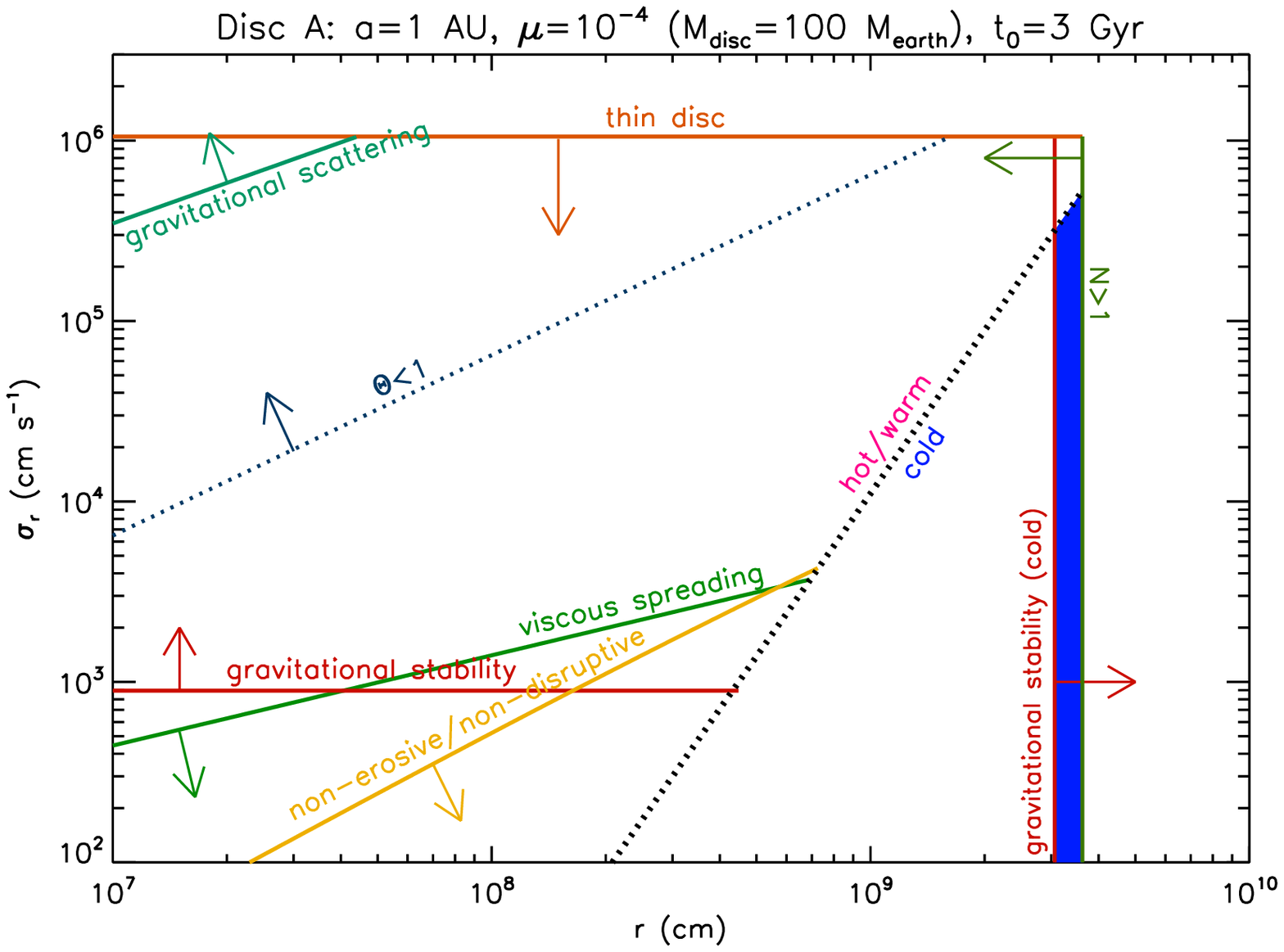}
\includegraphics[width=0.8\columnwidth]{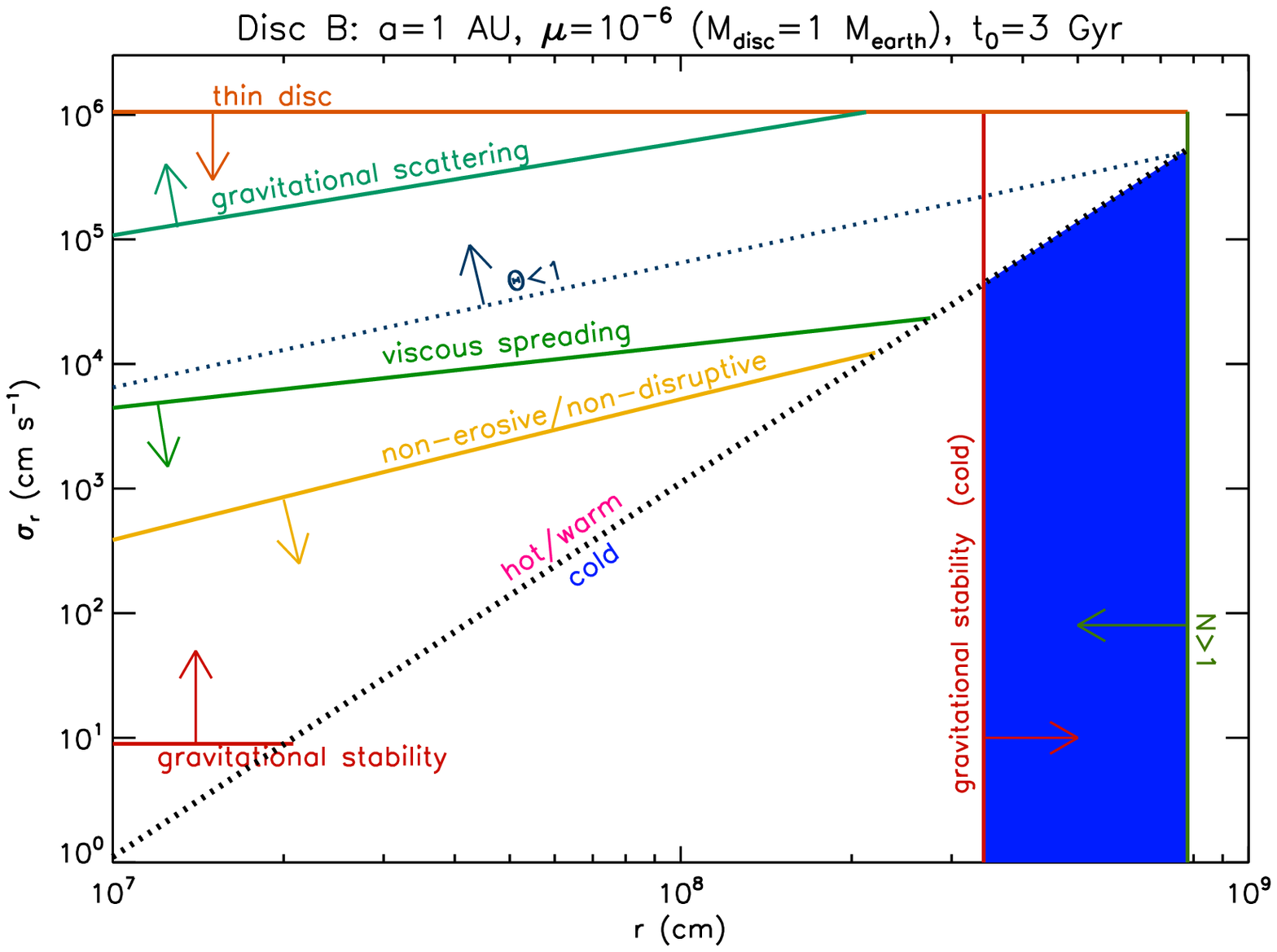}
\end{center}
\caption{Allowed values of planetesimal velocity dispersion and radius for
  discs with age $t_0 = 3\Gyr$, semi-major axis $a=1\au$, and dimensionless
  mass $\mu = 10^{-4}$ ($M_{\rm disc} \simeq 100$ $M_\oplus$) and $\mu =
  10^{-6}$ ($M_{\rm disc} \simeq M_\oplus$), i.e., discs A and B of Table
  \ref{tab:disc}.  The blue shaded region denotes allowed cold discs.  There
  are no allowed hot or warm discs.  The various lines represent conditions
  for: gravitational stability (equations [\ref{eq:hillstabb}] and
  [\ref{eq:tt}]), gravitational scattering time exceeds disc age (equation
  [\ref{eq:tgrav}]), thin disc (equation [\ref{eq:thin}]), $N>1$ (equation
  [\ref{eq:number}]), collisions are not erosive/disruptive (equations
  [\ref{eq:hotdiscs_type}] and [\ref{eq:qast}]), and viscous spreading time
  exceeds disc age (equation [\ref{eq:viscous}]). The arrows attached to each
  line indicate the region in which long-lived discs could exist. The dividing
  line between hot/warm and cold discs is given by equation (\ref{eq:hot}) and
  the dividing line between collision speeds greater than or less than the
  escape speed from the planetesimal surface ($\Theta<1$ or $\Theta>1$
  respectively) is given by equation (\ref{eq:thetaalt}).}
\label{fig:sample_discs1}
\end{figure}

\begin{figure}
\begin{center}
\includegraphics[width=0.8\columnwidth]{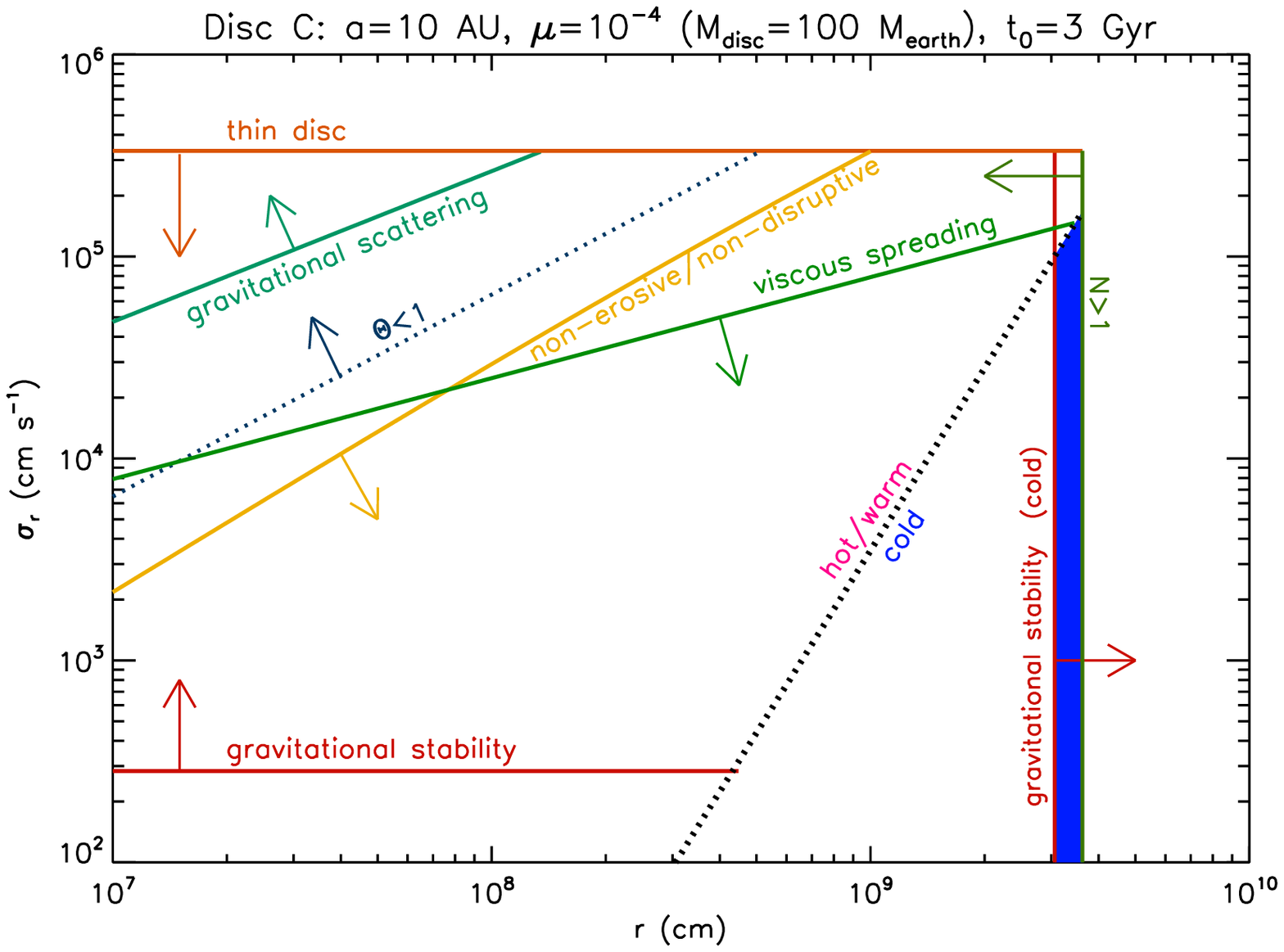}
\includegraphics[width=0.8\columnwidth]{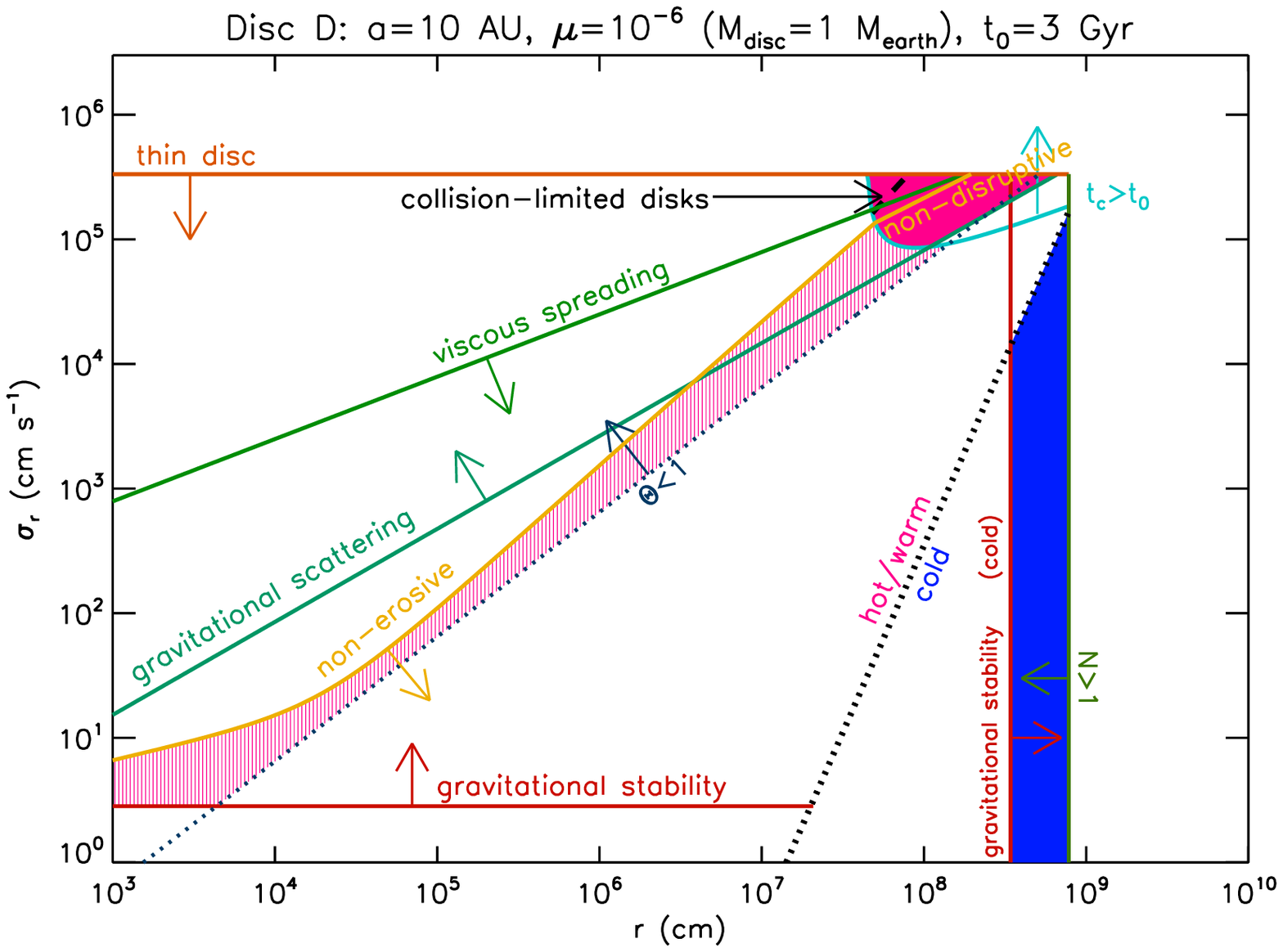}
\end{center}
\caption{Same as Figure \ref{fig:sample_discs1}, but for $a=10\au$, i.e.,
  discs C and D.  For disc D, the pink shaded regions denote allowed hot
  discs (solid color) and warm discs (vertical hatching).  An additional
  constraint for disc D is $t_c \gtrsim t_0$ (equations [\ref{eq:tcolla}] and
  [\ref{eq:tcollshear}]).  The dashed curve represents collision-limited discs
  (equation [\ref{eq:vcoll_limited}]).} 
\label{fig:sample_discs2}
\end{figure}

\begin{figure}[]
\begin{center}
\includegraphics[width=0.8\columnwidth]{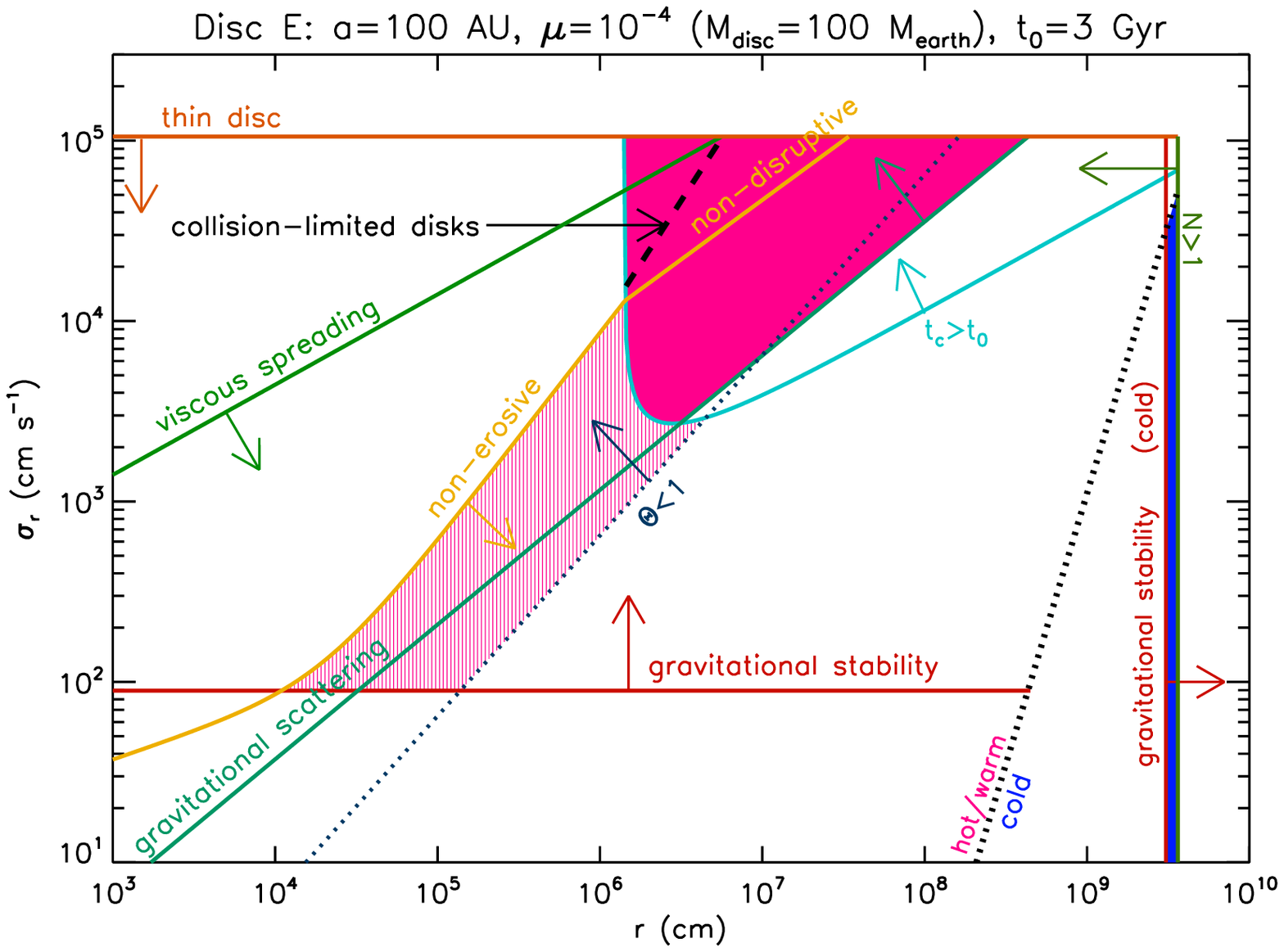}
\includegraphics[width=0.8\columnwidth]{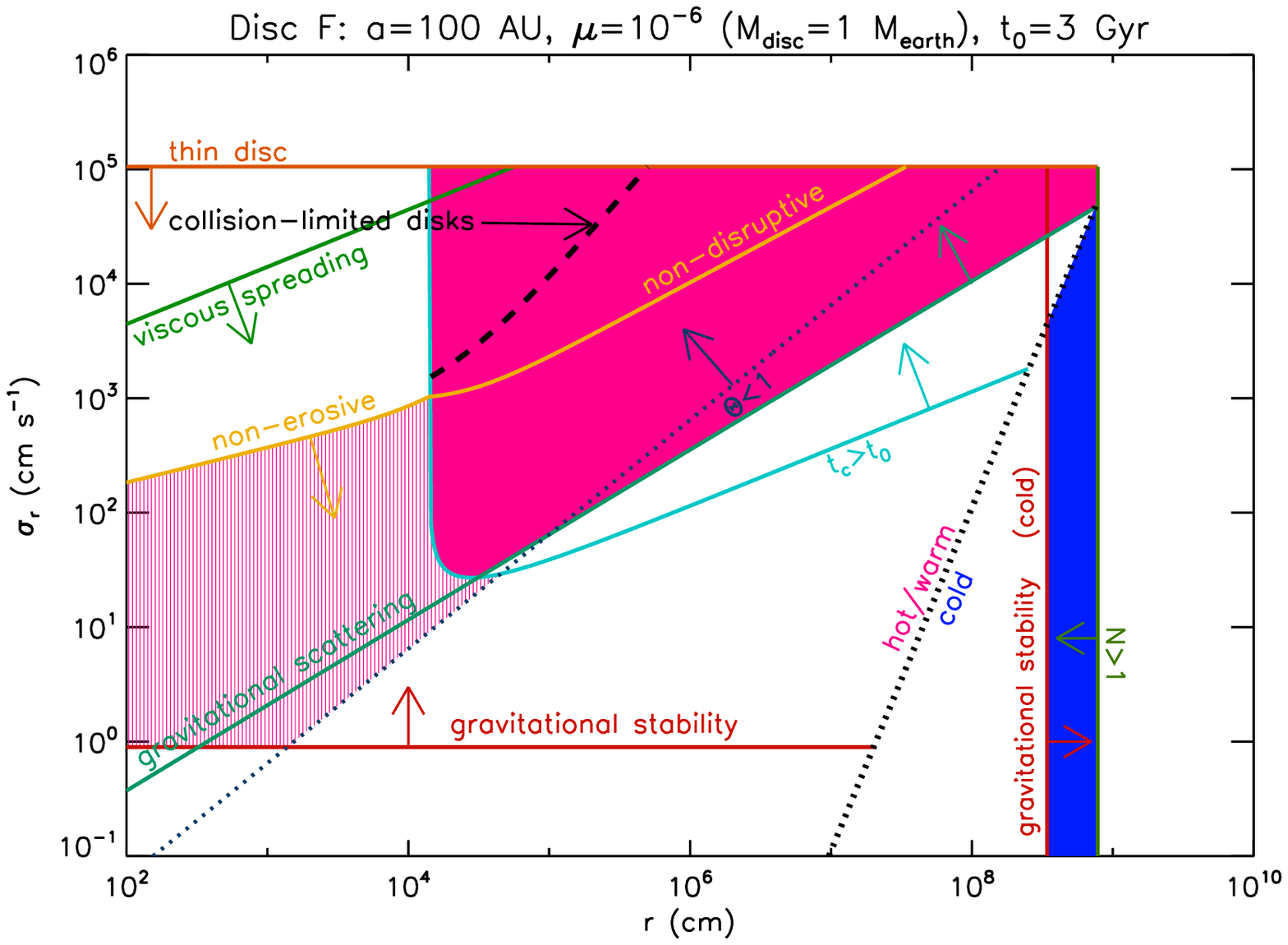}
\end{center}
\caption{Same as Figures \ref{fig:sample_discs1} and \ref{fig:sample_discs2},
  but for $a=100\au$, i.e., discs E and F.} 
\label{fig:sample_discs3}
\end{figure}

As described at the end of \S\ref{sec:simple}, we examine six possible
planetesimal discs (Table \ref{tab:disc}), with semi-major axes $a=1$, 10 and
100 AU, and dimensionless masses $\mu=10^{-4}$ and $10^{-6}$. The allowed
values of velocity dispersion $\sigma_r$ and planetesimal radius $r$ are shown
in Figures \ref{fig:sample_discs1} and \ref{fig:sample_discs2}. In all cases,
we assume that the disc age is $t_0=3$ Gyr and the planetesimal density is
$\rho_p=3$ $\gmcm$.

A. $a=1\au$, $\mu=10^{-4}$: Hot and warm discs cannot survive (see equations
[\ref{eq:hotdiscs_type}], [\ref{eq:exista}], [\ref{eq:existb}] and
[\ref{eq:warmdisc_mmax}]).  Cold discs {\em can} survive, but only for a
narrow range of planetesimal masses (the thin blue trapezoid in the upper
panel of Figure \ref{fig:sample_discs1}): there can be at most one or two
planetesimals per octave of semi-major axis, of mass $m^{\rm cold}_{\rm min}
\simeq 60 M_\oplus \simeq 0.2 M_{\rm Jupiter}$ (equation [\ref{eq:mmincold}]).
Such discs are rather similar to some of the many extrasolar planetary systems
already detected by radial-velocity variations in the host star.

B. $a=1\au$, $\mu=10^{-6}$: Hot and warm discs cannot survive. Cold discs can
have planetesimal masses in the range 0.1--$1 M_\oplus$ (equation
[\ref{eq:mmincold}]); the lower limit corresponds to about a dozen
planetesimals per octave of semi-major axis (equation [\ref{eq:nmaxa}]). These
discs may be detectable with space-based transit surveys and are reminiscent
of the terrestrial planets in our own solar system. 

C. $a=10\au$, $\mu=10^{-4}$:  Hot and warm discs cannot survive.  As in the
case of disc A, only one or two planetesimals (or planets) per octave of mass
$m^{\rm cold}_{\rm min} \simeq 60$ $M_\oplus$ can survive in a cold disc. We
suggest in \S\ref{subsect:classify_microlensing} that some discs of this type
may be detectable by gravitational microlensing. 

D. $a=10\au$, $\mu=10^{-6}$: Hot, warm, and cold discs can all survive.  The
hot discs may contain up to about 5000 planetesimals per octave with minimum
masses of $m^{\rm hot}_{\rm min} \approx 10^{24}$ g, about the mass of Ceres.
The dynamical constraints on cold discs are the same as for disc B; such
discs are not detectable with current or planned transit surveys because the
probability of transits is too small and the orbital period is too large but
could be detected by targeted searches for gravitational microlensing (see
Figure \ref{fig:classify}).  A wide range of warm discs is possible, with
at least 200 planetesimals per octave and masses at most
0.4 times that of the Moon. A more typical warm disc might have $5\times 10^5$
planetesimals per octave, of radius $100\km$, with velocity dispersion
$\sigma_r\simeq 0.1\kms$ and a collision time of $0.6\Gyr$. 

E. $a=100\au$, $\mu=10^{-4}$: Hot, warm, and cold discs can all survive. As in
the case of discs A and C, the cold discs can have only one or two
planetesimals per octave. Hot discs can have up to $2 \times 10^{10}$ objects
per octave with masses of at least $4 \times 10^{19}$ g.  Warm discs contain
at least $5 \times 10^8$ objects per octave with maximum masses of $10^{21}$ g
($r \simeq 40$ km; approaching the largest sizes of comet nuclei). 

F. $a=100\au$, $\mu=10^{-6}$:  Hot discs have at most $2 \times 10^{14}$
objects per octave with $m^{\rm hot}_{\rm min} \simeq 4 \times 10^{13}$ g.
Warm discs contain objects with $m \lesssim 10^{15}$ g ($r \simeq 400$ m), of
which there are at least $5 \times 10^{12}$ per octave.  The constraints on
cold discs are the same as for discs B and D. 

It is worthwhile to emphasize that Figures
\ref{fig:sample_discs1}--\ref{fig:sample_discs3} do not show how a
planetesimal disc of a given initial mass will evolve, but rather whether a
disc with a current mass $M_{\rm disc}$ can survive in approximately its
current state for $3\Gyr$; the `allowed' regions in these Figures can be
interpreted as the allowed regions for discs with an age $t_0=3\Gyr$ on the
assumption that it is unlikely to find objects in states that evolve on a
timescale much less than their age. 

\subsection{Collision-limited discs}
 
Collision-limited discs are shown by dashed curves in Figures
\ref{fig:sample_discs1} and \ref{fig:sample_discs2} for discs D, E and F
(equation [\ref{eq:vcoll_limited}]).  The allowed ranges of masses are
$10^{24}\gm \lesssim m_{\rm max} \lesssim 8 \times 10^{24} \gm$ (disc D), $4
\times 10^{19}\gm \lesssim m_{\rm max} \lesssim 3 \times 10^{21}\gm$ (disc E)
and $4 \times 10^{13}\gm \lesssim m_{\rm max} \lesssim 4 \times 10^{18}\gm$
(disc F).

\section{Detection techniques}
\label{sect:detection}

\subsection{Radial velocity measurements}
\label{subsect:rv}

The stellar wobble or reflex radial velocity induced by an
edge-on disc containing $N$ planetesimals of mass $m$ is approximately
\be
\begin{split}
v_{\rm wobble} &\simeq \sqrt{\frac{GM_\odot N}{a}}\frac{m}{M_\odot} \\
&= \pi f_m \mu \sqrt{\frac{GM_\odot}{aN}}\\
&= 3.0 \mbox{ m s}^{-1} ~f_m N^{-1/2} \left(\frac{a}{10\au} \right)^{-1/2}
\left(\frac{\mu}{10^{-4}} \right).  \\ 
\end{split}
\label{eq:wobble}
\ee

With current technology we can detect reflex velocities as small as $v_{\rm
  wobble}\sim 1\,\hbox{m s}^{-1}$ with orbital periods as long as $\sim
10\yr$, corresponding to $a \simeq 4.6\au$. For hot discs with ages of several Gyr,
equations (\ref{eq:existb}) and (\ref{eq:wobble}) imply
$v_{\rm wobble} \lesssim 0.1\,\hbox{m s}^{-1}(a/10\au)^{5/2}$, too small to be
detectable. 

Cold discs are detectable if the semi-major axis is
small: combining equation (\ref{eq:nmaxa}) with the second of the equations
above, we have 
\be
v_{\rm wobble} > 2.3 \mbox{ m s}^{-1} ~f_m^{1/2} \left(\frac{a}{10\au}
\right)^{-1/2} \left(\frac{\mu}{10^{-4}} \right)^{1.22}. 
\ee
Thus {\it all\/} long-lived cold discs with $\mu \gtrsim 10^{-4}$ and
$a\lesssim 10\au$ can be detected by current radial-velocity surveys; this of
course is because gravitational stability requires that they have only a few
large planets.   

\subsection{Transits}
\label{subsect:transit}

Space-based transit surveys such as NASA's {\it Kepler\/} mission are capable
of detecting photometric variations as small as $\sim 10^{-5}$, corresponding
to the transit of a planetesimal of radius $2000 \km=0.35R_\oplus$. Thus
edge-on discs containing lunar-mass planetesimals may be detectable by transit
surveys. Reliable detection of transits of such small objects requires that
the stellar variability is negligible; this is likely to be true for at least
some stars as the solar variability on the hourly timescales relevant to
transit detection is only a few times $10^{-5}$ \citep{jen02}.

Other criteria for detectability of edge-on planetesimal discs by transits
include the following: (i) The orbital period must be less than a year or so,
so that several transits of a given object can be detected in a mission of 
reasonable duration. (ii) There must not be too few
transits, that is, at least one planetesimal in the disc must transit the
star. If the disc is nearly edge-on and the characteristic thickness $h$
(equation [\ref{eq:nz}]) is small compared to the stellar radius $R_\star$
then most 
planetesimals transit the stellar disc in the course of an orbit, while if
$h\gg R_\star$ the fraction of transiting planetesimals is
$(2/\pi)^{1/2}R_\star/h$, so the expected number of transiting planetesimals
is roughly
\be
N_t=N\,\mbox{min}\left[1,\left(2\over\pi\right)^{1/2}{R_\star\over h}\right].
\label{eq:transitcond}
\ee
(iii) There must not be {\em too many} transits: if multiple planetesimals are
transiting the disc at any one time, the fluctuations in stellar flux will be
difficult to distinguish from normal stellar variability. The average number of
planetesimals in transit at a given time is $NR_\star/(\pi a)$ if $h\ll
R_\star$ and $NR_\star^2/(\sqrt{8\pi}ha)$ if $h\gg R_\star$ so the average
number of planetesimals in transit at any instant is 
\be
n_t=N\,\mbox{min}\left({R_\star\over\pi a},{R_\star^2\over\sqrt{8\pi}ha}\right).
\label{eq:transitcondb}
\ee

\subsection{Microlensing}
\label{subsect:microlensing}

\subsubsection{Microlensing by individual planetesimals}

The classical lensing equation is \citep[e.g.,][]{sch92}
\be
\alpha d_\star = \frac{d_\star L}{d} - \frac{2 R_s d_{{\rm l}\star}}{L},
\ee
where $R_s = 2Gm/c^2$ is the Schwarzschild radius of the lens and $L$ is the
projected separation between the light ray and the lens in the lens plane.
The distance to the source and lens and the separation between them are
denoted by $d_\star$, $d$ and $d_{{\rm l}\star}=d_\star-d$, respectively.  The quantity
$\alpha$ is the subtended angle between the lines of sight to the lens and the
source.  The Einstein radius is defined by the value of $L$ when 
$\alpha=0$ (i.e., lens and source are aligned),
\be
r_{\rm E} = \frac{2}{c} \sqrt{Gmd_\star \zeta \left(1-\zeta \right)}
\label{eq:ein}
\ee
where $\zeta\equiv d/d_\star$.

A planetesimal can be far from or near to its parent star, where `far' and
`near' are defined with respect to the stellar Einstein radius, 
\be 
R_{\rm  E}=4.0\au\left[{M_\star\over M_\odot}{d_\star\over
    8\kpc}{\zeta(1-\zeta)\over 0.25}\right]^{1/2},
\label{eq:einstein_star}
\ee
with $M_\star$ being the stellar mass.  

The magnification of the total flux from the 
source is
\be
{\cal A}={u^2+2\over u\sqrt{u^2+4}},
\ee
where $u \equiv \alpha d /r_{\rm E}$.  In our simple treatment, we assume that
a planetesimal can produce a detectable lensing event of non-negligible
magnification when the distance $\alpha d$ between the lines of sight to the
planetesimal and the source star, measured in the lens plane, is less than the
Einstein radius; this corresponds to $u=1$ or amplification ${\cal
  A}=3/\surd{5} \simeq 1.34$.

We must check that the planetesimal radius is small compared to the
Einstein radius to ensure that the magnified light curve is not blocked
\citep{agol02}. We have\invisible{actual value 0.6595}
\begin{eqnarray}
{r\over r_{\rm E}} &=& \left[3c^2\over 16\pi G\rho_prd_\star\zeta(1-\zeta)\right]^{1/2}
  \nonumber \\
 &=& 0.66\left[{r\over1\km}{\rho_p \over 3\gmcm}{d_\star \over 8\kpc}{\zeta\left(1-\zeta\right)\over 0.25}\right]^{-1/2},
\end{eqnarray}
so obscuration by the planetesimals is unimportant if they are much larger
than 1 km. 

In addition, we require that the stellar (source) radius
as projected on the lens plane -- equal to the stellar radius $R_\star$
multiplied by $\zeta$ -- cannot be much larger than the
Einstein radius, to ensure that the magnified light curve is not smeared
out. We have\invisible{actual number 7.26} 
\begin{eqnarray}
{R_\star \zeta\over r_{\rm E}} &=& \left({3c^2\ R_\star^2\over 16\pi
    G\rho_pr^3d_\star}{\zeta\over 1-\zeta}\right)^{1/2} \nonumber \\ 
 &=& 7.3 \left({R_\star\over R_\odot}\right) \left(r \over 1000\km
 \right)^{-3/2} \left({\rho_p \over 3\gmcm}{d_\star \over 8\kpc}{1-\zeta \over
     \zeta}\right)^{-1/2}. 
\label{eq:size_cutoff}
\end{eqnarray}
We have parametrized $R_\star$ in terms of the solar radius since this is the
typical size of the source star in existing planetary microlensing events (at
$8\kpc$ the corresponding angular size is $0.6\,\mu\hbox{as}$).  With the
nominal parameters and $\zeta=0.5$, $R_\star \zeta /r_{\rm E}<1$ only for
planetesimal radii $r>3750\km$, corresponding to mass $m>0.11M_\oplus$
(\citealt{pac96} gives a similar estimate, $0.07M_\oplus$). This limit is
conservative because the magnification of extended sources remains substantial
when the projected source radius is as large as several times the Einstein
radius -- for example, a uniform source whose centre is separated from the lens
by one Einstein radius is magnified {\em more} than a point source so long as
$R_\star \zeta /r_{\rm E}<2.17$ \citep{gou94,wm94}. Thus, microlensing searches
are likely to be sensitive to planetesimals as small as $\sim
10^{-1.5}M_\oplus$ or a few times the mass of the Moon (but see \citealt{hk09})\footnote{Events associated with lower
amplifications have larger microlensing cross sections, i.e., $\pi (\phi
r_{\rm E})^2$ where $\phi > 1$.  Planetesimals with Einstein radii smaller
than the projected size of the source star may contribute appreciably to the
expected number of events per planetesimal disc crossing, because the range of
masses involved in microlensing now extends down to much lower values.}.

If the transverse velocity of the lens relative to the
source is $v_\perp$, the characteristic duration of the event is
\be
t_{\rm E,d} \sim  \frac{2r_{\rm E}}{v_\perp} = 16\,\hbox{min} \left({v_\perp \over
  100\kms}\right)^{-1} \left(r\over 1000\km\right)^{3/2} \left[{\rho_p\over3\gmcm}{d_\star\over8\kpc}{\zeta(1-\zeta)\over 0.25} \right]^{1/2}.
\label{eq:duration} 
\ee
In most cases, the transverse velocity is dominated by the apparent angular
speed of the source star relative to the host star of the planetesimal, rather
than the motion of the planetesimal around its host star.

If we assume that the surface density of the disc is uniform over a circle of
radius $a$, the probability of lensing at any given moment for a star whose
image lies within the disc (i.e., the optical depth) is\invisible{actual value: 5.11e-5}
\be
\begin{split}
\tau_{\rm lens} &\simeq N \left(\frac{r_{\rm E}}{a}\right)^2\\
&=  5\times 10^{-5}f_m \left( \frac{\mu}{10^{-4}} {d_\star\over8\kpc}\right) \left(a \over 10\au \right)^{-2}{\zeta(1-\zeta)\over 0.25}.\\
\end{split}
\label{eq:tau_lens}
\ee
The optical depth of a planetesimal disc of a given size $a$ and
mass $M_{\rm disc}=\pi f_m\Sigma a^2$ is independent of the mass of the
individual planetesimals.  Thus (for example) the optical depth for a
disc composed of 100 Earth-mass planets ($M_{\rm disc}/M_\odot=3\times
10^{-4}$, $\mu=10^{-4}$) is the
same as the optical depth of a single 0.3 Jupiter-mass planet at the same
radius.

A related quantity is the probability that at least one lensing event by a
planetesimal will be seen at some time during the passage of the source star
near the host star of the planetesimal disc. If the impact parameter
associated with this passage is small compared to the size of the planetesimal
disc, this probability is $1-\exp(-\tilde{N}_{\rm lens})$, where\invisible{actual value
  9.10e-3}
\be
\begin{split}
\tilde{N}_{\rm lens} &= {4Nr_{\rm E}\over\pi a}\\
&= {8\over c}\left[{f_m\over\pi}G\Sigma d_\star N\zeta(1-\zeta)\right]^{1/2} \\
&= 9.1\times 10^{-3} \left({10\au\over a}\right) \left[f_m N{d_\star\over8\kpc}{\mu\over 10^{-4}}{\zeta(1-\zeta)\over 0.25}\right]^{1/2}.\\
\end{split}
\ee
Notice that the probability of observing an event goes up as the planetesimal
mass goes down, since $N\sim \Sigma a^2/m$ (although the duration of the event
is shorter). In this respect, a disc consisting of many small planetesimals may
actually be {\em easier\/} to detect than a single large planet.

The short duration (\ref{eq:duration}) of the events is one of the principal
challenges in reliably observing microlensing by planetesimals of an Earth
mass or less. To avoid being swamped by noise it is useful to focus on source
stars that are experiencing -- or have recently experienced -- microlensing
by an intervening star. Strong amplification by the host star of a
planetesimal disc requires that the impact parameter is less than the Einstein
radius of the host star, given by equation (\ref{eq:einstein_star}).  Since
many planetesimal discs may be substantially larger than $R_{\rm E}$ the
source star should be monitored for short-duration events for some time after
the amplification by the host star has returned to unity.  

\subsubsection{Other microlensing effects}

Planetesimal discs can produce other signals in microlensing surveys.
\cite{zm05} point out that if the source star for a microlensing event hosts a
debris disc, the mid/far-IR light curve will contain a component determined by
the surface-brightness profile of the thermal emission from the debris disc.
Similarly, the optical and near-IR light curve will contain a component from
the scattered light from the disc. In both cases we may expect that the light
curve is no longer wavelength-independent. These effects are challenging to
detect since (i) many debris discs are much larger than the stellar Einstein
radius $R_{\rm E} \simeq 4\au$ (equation [\ref{eq:einstein_star}]), so the
maximum magnification is only $\sim (R_{\rm E}/a)^2$; (ii) accurate mid/far-IR
photometry is exceedingly difficult, except from space; (iii) the fractional
flux of scattered light is small, typically $10^{-3}$ to $10^{-5}$ in observed
debris discs (e.g., \citealt{wy08}).

Other signals may arise if the lens star hosts a planetesimal disc. The
overall mass distribution in the disc will contribute to the magnification and
thereby distort the microlensing light curve, but this distortion will be
difficult to detect because the magnification due to the disc will only be of
order $\mu R_{\rm E}/a$ where $\mu$ is defined in equation (\ref{eq:dimen});
we have assumed that the disc is not far from face-on and that its semi-major
axis $a$ is larger than the Einstein radius of the host star (see
\citealt{hun09} for a discussion of lensing by edge-on discs).  A potentially
more sensitive probe is high-magnification events. The gravitational field
from distant stellar companions or other external mass distributions can
produce a characteristic double-peak structure near the point of maximum
magnification; for example, \cite{kim08} estimate that typical
high-magnification events can detect stellar companions with mass $m$ and
separation $\Delta d$ such that $m/M_\odot \gtrsim (\Delta d/100\au)^2$.
Unfortunately, these events are insensitive to distant discs in most cases, for the
following reason: the deflection angle from a surface mass density
distribution $\Sigma_{\rm proj}({\bf x})$ on the sky plane is \citep{sch92}
\be
{\bf \alpha}({\bf x})=\int {4G\Sigma_{\rm proj}({\bf x}^\prime)\over c^2}
{{\bf x}-{\bf x}^\prime\over \vert{\bf x}-{\bf x}^\prime\vert^2} d^2{\bf x}^\prime.
\ee
which is proportional to the gravitational field from a cylindrical mass
distribution with density $\rho(x_1,x_2,x_3)\propto\Sigma_{\rm
  proj}(x_1,x_2)$. The deflection angle for light rays passing inside an
inclined ring of material is therefore proportional to the gravitational field
inside an elliptical, cylindrical shell, which is zero from Newton's theorem.
Only discs in which $\Sigma_{\rm proj}$ is non-zero near the host star (e.g.,
nearly edge-on discs with a significant thickness) will affect the light curve
near the peak magnification. 

\subsection{Infrared emission due to dust generated from collisions}
\label{subsect:fir}

\begin{figure}
\begin{center}
\includegraphics[width=0.8\columnwidth]{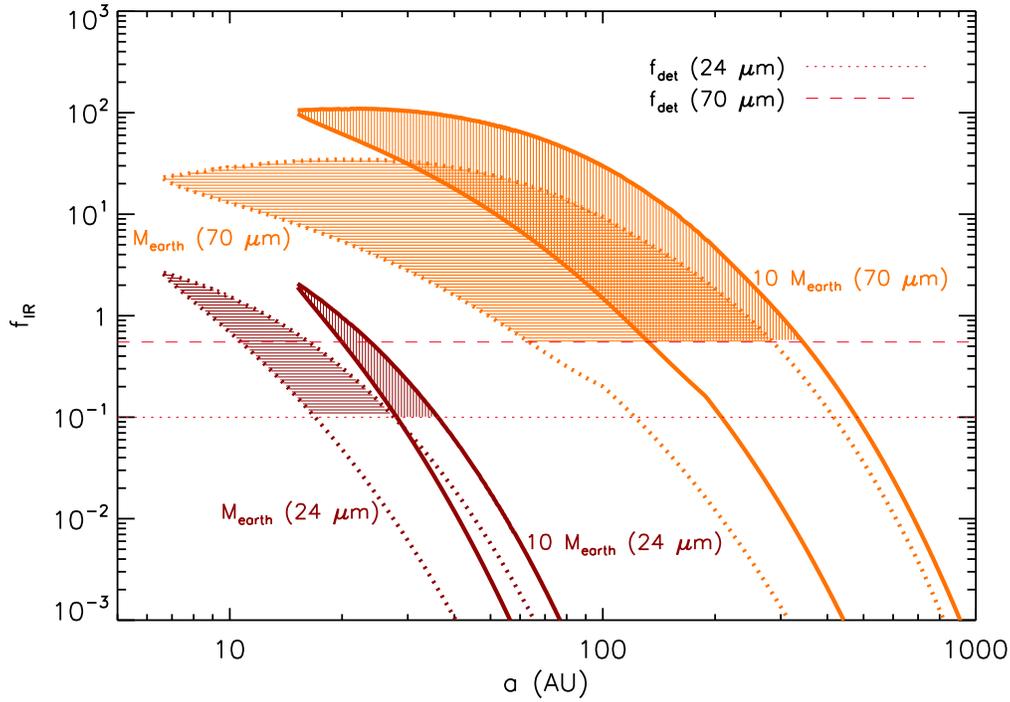}
\end{center}
\caption{Allowed values of the infrared excess, $f_{\rm IR}(\lambda)$, for hot discs
  around solar-type stars with masses $M_{\rm disc} = M_\oplus$ and 10
  $M_\oplus$.  The shaded regions denote hot discs that are detectable via
  their IR excesses; the assumed detection thresholds $f_{\rm det}$ at
  $\lambda = 24$ and $70\,\mu$m are shown as horizontal lines.} 
\label{fig:fir_allowed}
\end{figure}

Most extrasolar detections of planetesimal discs are based on measurements of
IR excesses, i.e., the presence of a debris disc \citep{wy08} in which a
steady supply of dust is generated by planetesimal collisions; the dust is
heated by the host star and the heated dust generates IR emission. At a given
wavelength, the IR excess, $f_{\rm IR}$, is the dust luminosity measured
relative to the stellar luminosity\footnote{Confusingly, $f$ is also used by
  many authors to denote the bolometric luminosity of the disc relative to the
  star. For particles with absorption efficiency $Q_a=1$ (equation
  [\ref{eq:mie}]), this ratio is equal to the geometrical optical depth
  $\tau_p$ defined in equation (\ref{eq:taupdef}).}. In this subsection we
estimate the IR excess due to planetesimal collisions in discs. The discs
considered here differ from the collision-limited discs of \S\ref{sec:col} in
that the collision time exceeds the age of the discs, $t_c\gtrsim t_0$.

If the dust grains are treated as gray bodies, their equilibrium temperature
is
\be
T_{\rm dust} = \left({\cal L}_\star\over 16\pi a^2\sigma_{\rm SB}\right)^{1/4}
\!\!= 279\, \mbox{K} ~\left(\frac{\au}{a}\right)^{1/2} 
\left(\frac{{\cal L}_\star}{{\cal L}_\odot}\right)^{1/4}\!\!,
\label{eq:dust_temp}
\ee
where $\sigma_{\rm SB}$ is the Stefan--Boltzmann constant. The peak wavelength of the
black-body spectrum $\lambda B_\lambda(\lambda,T)$ at
this temperature is $\lambda_{\rm max} =
13\,\mu\mbox{m}\,(a/\au)^{1/2}({\cal L}/{\cal
  L}_\odot)^{-1/4}$.
The IR excess is then
\be
f_{\rm IR}(\lambda) = \frac{B_\lambda\left(\lambda,T_{\rm
      dust}\right)}{B_\lambda\left(\lambda,T_\star\right)}
\int ~Q_a\left(\frac{r}{R_\star}\right)^2 ~dN_{\rm dust},
\label{eq:fir}
\ee
where $T_\star$ and $R_\star$ are the stellar temperature and radius, $Q_a$ is
the absorption efficiency, and $dN_{\rm dust}(r)$ is the number of dust
particles as a function of their radius $r$. As discussed at the end of
\S\ref{sec:dust}, the absorption efficiency can be approximated as
\be
  Q_a=\hbox{min}\,\{1,X\}\quad\hbox{where}\quad  X\equiv {2\pi r\over\lambda},
\label{eq:mie}
\ee
so if the number of dust particles is a power law in radius, 
\be
dN_{\rm dust}=K r^{-q}dr,
\label{eq:dndustdr}
\ee
we have
\be
f_{\rm IR}(\lambda) = \frac{B_\lambda\left(\lambda,T_{\rm
      dust}\right)}{B_\lambda\left(\lambda,T_\star\right)}
{K\over (4-q)(q-3)R_\star^2}
\left(\lambda\over 2\pi\right)^{3-q}.
\label{eq:fire}
\ee

The radius exponent $q$ is related to the mass exponent $p$ defined in
equation (\ref{eq:massdist}) by $q=3p-2$. For small particles such as dust, we
expect that the specific kinetic energy required for disruption, $Q_D^\ast$,
is independent of mass (cf.\ \S\ref{sec:coleff}). In this case $p=11/6$
(equation (\ref{eq:massexp})] so $q=7/2$, and we shall use this value in evaluating
equation (\ref{eq:fire}) numerically. Note that equation (\ref{eq:fire}) is only
valid if $3 \le q \le 4$ and if the minimum grain size in the distribution is
much smaller than $\lambda/2\pi$.

The physical processes governing the effects of collisions are outlined in
\S\ref{sec:coleff}. In a steady state, the rate of dust mass production 
in a monodisperse planetesimal disc of the kind we are considering is
\be
\Phi_m \approx \frac{mN}{t_c},
\label{eq:massflux}
\ee
where as usual $m$ and $N$ are the mass and total
number of planetesimals and $t_c$ is the collision time, given by equation
(\ref{eq:tcolla}).  To estimate the corresponding dust mass, we use the 
conservation of mass flux. 

First we generalize the collision time (\ref{eq:tcolla}) to the case where
particles of radius $r_1$ are colliding with particles of radius
$r_2$. We have
\be
t_c^{-1}\left(r_1\right)=2^{3/2}\pi^{1/2}f_1\sigma_r\int_{r_{\rm min}}^{r_{\rm max}}(r_1+r_2)^2
  \frac{dn_0\left(r_2\right)}{dr_2} \,dr_2;
\ee
here we have assumed that self-gravity is negligible ($\Theta\ll1$). If the
number density is a power law in radius, $dn_0(r)/dr \propto r^{-q}$, and the
integral is dominated by projectiles with radii $r_2$ much less than the target radius
$r_1$, we have 
\be
t_c^{-1}\left(r_1\right)={2^{3/2}\pi^{1/2}f_1\over q-1}
\frac{dn_0\left(r_1\right)}{dr_1} \sigma_r r_1^3 \theta^{q-1} 
\ee
where $\theta\equiv r_1/r_{\rm min}\gg1$, i.e., $r_1 \gg r_{\rm min}$.  If we
define $r_{\rm min}$ to be the 
minimum projectile radius that will disrupt a grain of radius $r_1$,
then the mass flux in the disc is roughly
\be
\Phi_m(r_1) \approx {m(r_1)\over t_c(r_1)}\left(dN_{\rm dust}\over d\log
  r\right)_{r_1},
\label{eq:massfluxdust}
\ee
where $m(r_1)=\frac{4}{3}\pi\rho_pr_1^3$ is the mass of a grain of radius
$r_1$. The number density and the total number of grains are related by
equations (\ref{eq:rhosig}) and (\ref{eq:vdisp}),
\be
\left(dN_{\rm dust}\over d\log  r\right)_{r_1}=Kr_1^{1-q}=
2^{1/2}\pi^{3/2}f_m{i_0\over e_0}{\sigma_r a^2\over\Omega}r_1
\frac{dn_0\left(r_1\right)}{dr_1}. 
\label{eq:xx}
\ee

Equating equations (\ref{eq:massflux}) and (\ref{eq:massfluxdust}), using
equation (\ref{eq:xx}) to eliminate $dn_0(r_1)/dr_1$ in favor of $K$, and
assuming that the velocity dispersion, internal density, and radial
distribution of the dust and planetesimals are equal, we have 
\be
\theta^{5/2}K^2 \approx 10N^2r^5(1+f_2\Theta/f_1),
\label{eq:kdef}
\ee
which relates the properties of the dust distribution on the left side
of the equation to those of the parent planetesimals on the right. Although we 
have assumed $q=7/2$ for the dust particles (i.e., the particles that
dominate the IR emission), which is equivalent to the assumption that the
specific dispersion energy $Q_D^\ast$ is independent of mass for these
particles, this derivation does not require any assumption about the 
dependence of $Q_D^\ast$ on mass for larger bodies -- the argument relates
the mass flux from the destruction of equal-mass planetesimals (equation 
[\ref{eq:massflux}]) to the mass flux in dust (equation [\ref{eq:massfluxdust}]) and
since mass flux is conserved the properties of intermediate-mass bodies are
irrelevant. The value of $\theta$ is estimated
from the specific dispersion energy $Q^\ast_D$; following the discussion after
equation (\ref{eq:qast}) a typical value is $Q^\ast_D= 10^7\,\hbox{erg
  g}^{-1}$ and we equate $m(r_1)Q^\ast_D$ to $\half m(r_{\rm min})\sigma_r^2$ to obtain
\be
\theta=28\, e_0^{2/3}\left({a \over 10\au}
{Q^\ast_D \over 10^7\,\hbox{erg g}^{-1}}\right)^{-1/3}.
\ee
The approximations that we have made to derive equation (\ref{eq:kdef})
are invalid unless $\theta\gg1$.

Equation (\ref{eq:kdef}) determines the normalization $K$ of the dust
distribution, which is substituted into equation (\ref{eq:fire}) to determine
the IR excess. This derivation is for hot discs, and implicitly assumes a
monodisperse disc in which collisions slowly feed a population of smaller
debris. These assumptions are only valid for discs in which the collision time
exceeds the age, $t_c\gtrsim t_0$. Once $t_c\sim t_0$ the appropriate
model is a collision-limited disc (\S\ref{subsect:coll_limited}), and the
factor $t_c$ in equation (\ref{eq:massflux}) should be replaced by $t_0$. The
analogue to equation (\ref{eq:kdef}) is then
\be
\theta^{5/2}K^2 \approx \frac{15}{32\Omega t_0}\frac{f_m}{f_1}\frac{M_{\rm
    disc}a^2}{\rho_p}. 
\label{eq:kdefa}
\ee

As discussed in \S\ref{subsect:dynamics_collisions}, collisions also occur in
warm discs, but in this case the collisions do not disrupt the planetesimals
and a collisional cascade is not established. We evaluate the IR emission
properties of warm discs in \S\ref{subsect:classify} below by assuming that most of
the emission comes from the planetesimals themselves.

For calibration-limited detections -- in which the limiting factor is the
accuracy of the extrapolation of the photospheric flux to long
wavelengths\footnote{See \S2.4 of \cite{wy08} for a discussion of calibration-
  vs. sensitivity-limited surveys.} -- the minimum detectable flux produced
by the dust, normalized by the stellar flux, is $f_{\rm det}$.  For
illustration, we set $f_{\rm det} = 0.1$ ($\lambda=24\,\mu$m) and 0.55
($\lambda=70\,\mu$m), similar to the limits in \cite{su06}.  We show examples
of hot planetesimal discs with detectable IR excesses in Figure
\ref{fig:fir_allowed}.  It is apparent that the IR excess is a poor diagnostic
for the disc mass $M_{\rm disc}$: the infrared flux from discs of a given mass
and semi-major axis in Figure \ref{fig:fir_allowed} can vary by more than an
order of magnitude.

\subsection{Probing disc mass and size}
\label{subsect:classify}

\begin{figure}
\begin{center}
\includegraphics[width=0.8\columnwidth]{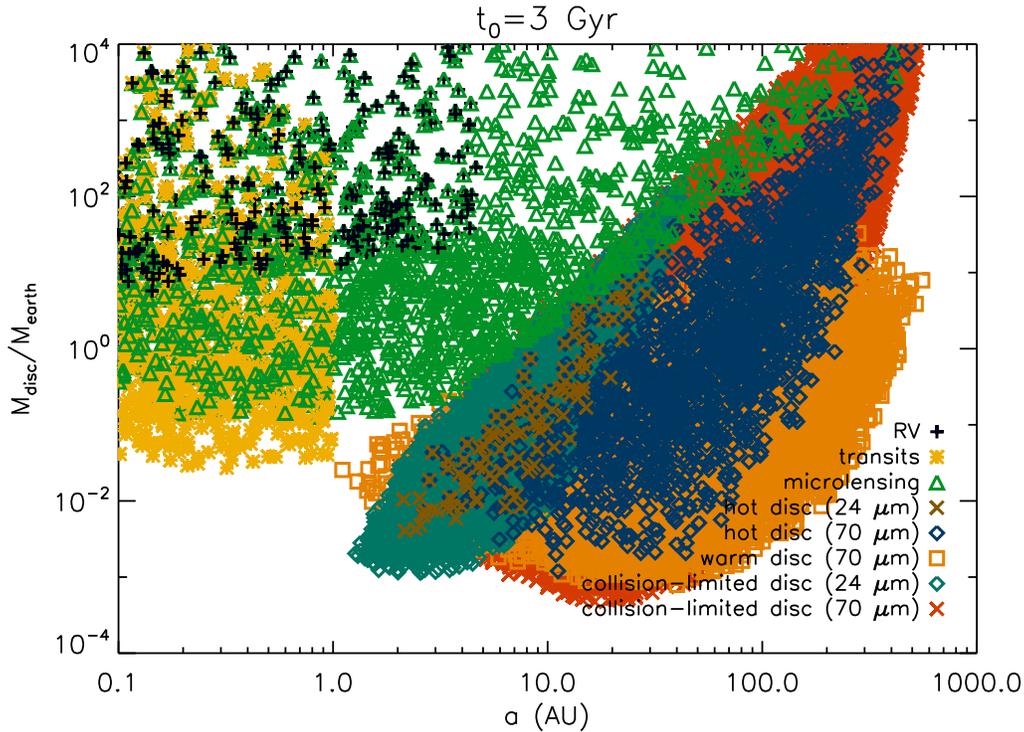}
\end{center}
\caption{Different detection techniques probe different ranges of planetesimal
  disc mass and semi-major axis.  Each of the four disc parameters (mass,
  semi-major axis, planetesimal radius, and velocity dispersion) is
  randomly generated and the various detection criteria are checked (see
  text). Only discs that survive for $3\Gyr$ are shown. The detectability
  criteria are summarized in 
  \S\S\ref{subsect:classify_rv}--\ref{subsect:classify_warm}.}  
\label{fig:classify}
\end{figure}

We now ask what long-lived planetesimal discs are detectable by the methods we
have discussed in \S\S\ref{subsect:rv}--\ref{subsect:fir}.  To efficiently
explore the four-parameter space of disc mass and semi-major axis,
planetesimal radius, and velocity dispersion ($M_{\rm disc}, a, r, \sigma_r
$), we randomly generate $3\times10^6$ discs, uniformly sampled on
logarithmic scales: $10^{-4} \le M_{\rm disc}/M_\oplus \le 10^4$, $0.1 \le
a/\au \le 1000$, $10^{-6} \le r/\mbox{\,cm} \le 10^{12}$ and $10^{-5} \le
\sigma_r/\mbox{\,cm s}^{-1} \le 10^7$.  We then ask whether each disc can survive for $t_0=3\Gyr$ and is
detectable by one or more methods using the detection thresholds described
below. Figure \ref{fig:classify} shows the detectable planetesimal discs as
projected onto the $M_{\rm disc}$--$a$ plane\footnote{Note that the density of
  generated points in Figure \ref{fig:classify} is generally lower at higher
  disc masses ($M_{\rm disc} \gtrsim 100 M_\oplus$), which is a surprising
  result since massive discs should be easier to detect. The low density comes
  about because the range of allowed planetesimal radii for cold discs becomes
  narrower for higher disc masses (see \S\ref{sec:sample} and Figures
  \ref{fig:sample_discs1}, \ref{fig:sample_discs2} and
  \ref{fig:sample_discs3}), and we are sampling $\log_{10} r$ uniformly.}.

\subsubsection{Radial velocities}
\label{subsect:classify_rv}

We consider a planetesimal disc to be detectable by this method if the orbital period
$2\pi(a^3/GM_\odot)^{1/2}$ is less than ten years and $v_{\rm wobble} \ge
v_{\rm det}$, where $v_{\rm wobble}$ is given by equation (\ref{eq:wobble})
and the detection threshold $v_{\rm det} = 1$ m s$^{-1}$. Black crosses in
Figure \ref{fig:classify} denote discs that survive for 3 Gyr and are detectable by this
method. The minimum detectable mass scales $\propto a^{1/2}$, as
expected. No warm or hot discs of age 3 Gyr were detectable by this method. 
Not surprisingly, planetesimal discs that are detectable by radial velocity
variations in the host star tend to be massive and contain a small number of
large bodies, i.e., planets; they resemble Disc A of \S\ref{sec:sample}. 

\subsubsection{Transits}
\label{subsect:classify_transit}

As described in \S\ref{subsect:transit}, planetesimals in nearly edge-on discs
can be detected transiting their parent
star if $r \gtrsim 0.35 R_\oplus$, and $a \lesssim 1\au$. We also require that
$N_t>1$ (equation [\ref{eq:transitcond}]) and $n_t<1$ (equation 
[\ref{eq:transitcondb}]). It could be argued that the condition $N_t>1$ is
unnecessarily stringent, since even if $N_t\ll1$ a fraction of discs with these
properties could be detected in a large transit survey. 

No warm or hot discs of age 3 Gyr were detectable via transits. Discs A and B of \S\ref{sec:sample} are detectable
via transits.

\subsubsection{Microlensing}
\label{subsect:classify_microlensing}

We consider a planetesimal disc to be detectable via microlensing if the
microlensing optical depth $\tau_{\rm lens}$ (equation [\ref{eq:tau_lens}])
exceeds $10^{-6}$ (for comparison, the measured microlensing optical depth
towards the Galactic bulge is 2--$3\times 10^{-6}$). Since $\tau_{\rm
  lens}$ is maximized when the planetesimal is halfway to the source, we adopt
$\zeta=d/d_\star=0.5$ for illustration.  We also assume a solar-type source
star at a distance $d_\star = 8\kpc$.  With these parameters, no warm or hot
discs are detectable by microlensing. In Figure \ref{fig:classify}, the cutoff
for $M_{\rm disc}\lesssim 0.1M_\oplus$ arises because the projected source
size becomes larger than the Einstein radius, while for $a \gtrsim 1\au$ we
have the detectable disc mass scaling $\propto a^2$ (equation
[\ref{eq:tau_lens}]). Discs A, B, and C of \S\ref{sec:sample} are detectable
by microlensing.

\subsubsection{Debris discs}
\label{subsect:classify_debris}

Debris discs are dynamically hot discs that produce a collisional cascade of
dust whose associated IR excess exceeds the detection threshold, $f_{\rm
  det}$.  We consider both hot and collision-limited discs.  As discussed in \S\ref{subsect:fir}, we take $f_{\rm det} = 0.1$ ($24\,\mu$m) and 0.55 ($70\,\mu$m).  In Figure \ref{fig:classify}, there is a minimum and maximum detectable disc mass at a given semi-major axis. This
feature appears to arise because the relaxation time $t_{\rm relax}$ (equation
[\ref{eq:trelax}]) at fixed disc mass and velocity dispersion has a minimum
near $\Theta=1$. There is also a cutoff at small semi-major axes, which arises
because the maximum allowable surface density for hot discs (equations [\ref{eq:exista}] and
[\ref{eq:existb}]) is a strongly increasing function of semi-major axis.  Discs
E and F of \S\ref{sec:sample} are detectable via their IR excesses at $70\,\mu$m.

It is remarkable that the discs detectable by radial-velocity/transit surveys
or microlensing do not overlap with those detectable from IR excess.  This
result is consistent with the observational findings of \cite{bei05},
\cite{bry06}, \cite{gfw06}, \cite{mm07}, and \cite{kospal09} that there is
little or no correlation between the occurrence of planets and debris discs.

This lack of overlap does not preclude the possibility that a single host star
may have planetesimals that are detectable by both methods, so long as the
planetesimal disc extends over several octaves in semi-major axis.
\cite{bei05} and \cite{kospal09} list six and ten planet-bearing stars with
debris discs, and planets have been imaged in the debris-disc systems HR 8799
\citep{mar08} and Fomalhaut \citep{kal08}.

\subsubsection{Warm discs}
\label{subsect:classify_warm}

In warm discs collisions may cause cratering of the planetesimals but do not
shatter them. Although cratering collisions produce significant amounts of
dust, they do not establish a collisional cascade of the kind described in
\S\ref{sec:cc}, so it is likely that IR emission is dominated by the
planetesimals themselves.  Based on this assumption we show the detectability
of warm discs from their IR emission in Figure \ref{fig:classify}. We again
adopt $f_{\rm det} = 0.55$ at $70\,\mu$m (we choose not to show detectable
warm discs at $24\,\mu$m so as not to over-crowd Figure \ref{fig:classify}).

In the Figure we see that there is again a minimum and maximum
detectable disc mass at a given semi-major axis.  This feature is now
associated with the constraints set by non-gravitational forces ($t_{\rm PR}
\gtrsim t_0$ and $\beta < 0.5$; see \S\ref{sec:dust}) and condition
(\ref{eq:hotdiscs_type}) for warm discs, respectively.

In principle, cold discs may also be detectable through the IR emission from
the planetesimals, but we found no such discs given our assumed detection
limits at $24\,\mu$m and $70\,\mu$m. 

\subsubsection{Undetectable discs}
\label{subsect:classify_others}

Many planetesimal discs that survive for 3 Gyr are not detectable using any of
the methods described in this section. 

\section{Discussion and summary}

\label{sec:disc}

\subsection{Can warm discs mimic debris discs?}
\label{subsect:warm_mimick}

\begin{figure}
\begin{center}
\includegraphics[width=0.8\columnwidth]{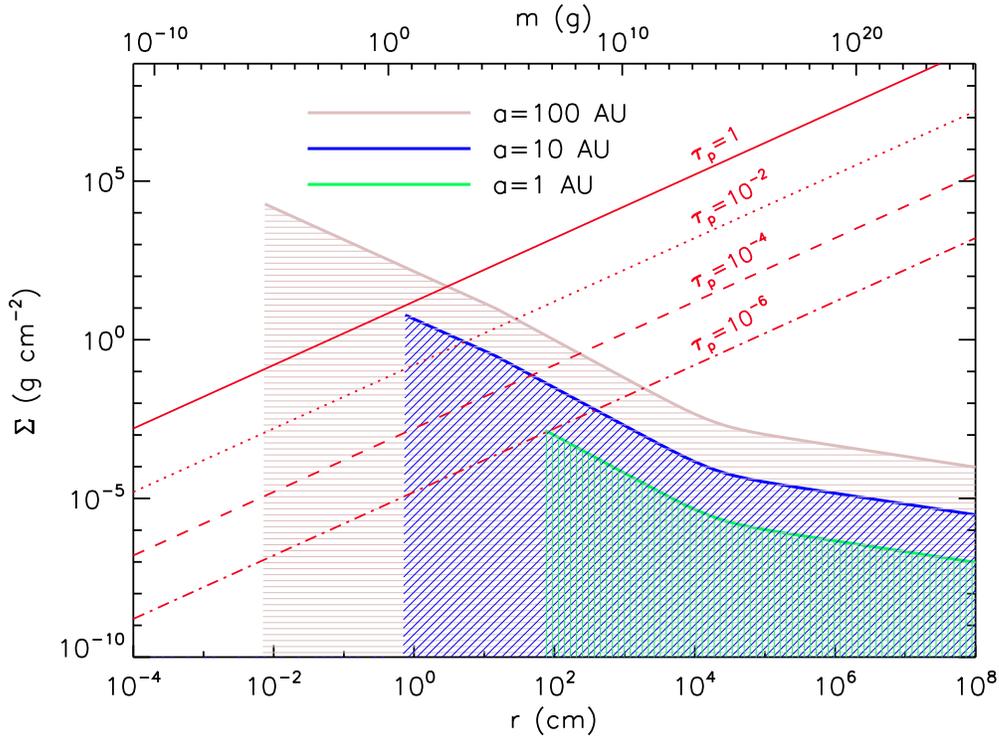}
\end{center}
\caption{Allowed surface density of warm discs as a function of the
  planetesimal radius according to equations (\ref{eq:warm_sigma}) and
  (\ref{eq:warm_pr}). The allowed region is shaded in different colours for
  different values of the semi-major axis $a=1$, 10 and 100 AU.  The disc age
  is assumed to be $3\Gyr$. Also shown are lines of constant optical
  depth $\tau_p$ (equation [\ref{eq:taupdef}]).}
\label{fig:warm}
\end{figure}

It is generally believed that the IR excesses around main-sequence stars older
than a few Myr are due to dust that is produced in collisions between large
solid bodies orbiting the star (hence the term `debris disc'). Direct
evidence that the emitting material is dust comes from several sources: (i)
Submillimeter observations of a handful of debris discs show that the
absorption efficiency $Q_a$ (equation [\ref{eq:fir}]) declines roughly as
$\lambda^{-1}$ for wavelengths $\gtrsim 100\,\mu$m \citep{de00,wa06,ba09},
suggesting grain sizes of a few tens of $\mu$m (equation [\ref{eq:mie}]).
(ii) \cite{chen06} obtained {\it Spitzer Space Telescope\/} infrared spectra
of 59 stars with IR excesses and found five with 10--$20\,\mu$m features that
imply the presence of micron-sized silicate grains. (iii) The polarization of
scattered light from the debris discs around $\beta$ Pictoris and AU
Microscopii is consistent with simple models of scattering by dust
\citep{gle91,gra07}.

Despite this evidence, it is instructive to consider the possibility that in
some stars the IR excess arises not from dust produced by a collisional
cascade but rather from a population of planetesimals with much larger
radii. The most likely candidates are warm planetesimal discs, 
in which the collision time is less than the disc age but the collision
velocities are too small to destroy the planetesimals over the lifetime of the
disc. To simplify the calculations, we consider the lowest possible radial
velocity dispersion for warm discs, which occurs when the Safronov number
$\Theta=1$ (cf.\ discs D, E, and F in Figures \ref{fig:sample_discs2} and
\ref{fig:sample_discs3}). Using equations
(\ref{eq:safronov}), (\ref{eq:tcolla}) and the third condition in equation
(\ref{eq:hotdiscs_type}), we obtain 
\begin{equation}
r \le \frac{1}{8\left(f_1 + f_2\right)} \left(\frac{Q^\ast_D}{G\Sigma \Omega t_0}\right).
\label{eq:warm_sigma}
\end{equation}
If $\Theta < 1$, then the preceding constraint becomes stronger, i.e., the
numerical coefficient in equation (\ref{eq:warm_sigma}) becomes larger. 

The minimum planetesimal size in warm discs is typically set by
Poynting--Robertson drag.  Using equations (\ref{eq:beta}) and (\ref{eq:tpr}),
the condition $t_{\rm PR} \gtrsim t_0$ yields 
\begin{equation}
r \gtrsim 0.7\hbox{\,cm}~{\cal P}_\phi \left(\frac{t_0}{3 \Gyr}\right)
\left(\frac{\rho_p}{3\gmcm} \right)^{-1} \left(\frac{a}{10\au} \right)^{-2}.
\label{eq:warm_pr}
\end{equation}

In Figure \ref{fig:warm} we show the constraints (\ref{eq:warm_sigma}) and
(\ref{eq:warm_pr}) for $a = 1,$ 10, and 100 AU; also shown are lines of
constant optical depth $\tau_p$ (equation [\ref{eq:taupdef}]). This optical
depth is equal to the ratio of the bolometric disc luminosity to the
bolometric stellar luminosity and hence provides a convenient measure of the
detectability of the disc. Known discs typically have $\tau_p\gtrsim 10^{-5}$
\citep{wy08}. We conclude from Figure \ref{fig:warm} that the IR emission from
so-called `debris discs' at $a=100\au$ could in some cases be coming from
planetesimals as large as $r \approx 10$ m ($m \approx 10^{10}$ g). A strong 
test of this possibility is that the emission spectrum from such a disc should
resemble a black-body spectrum, even at submm wavelengths (or a superposition
of black-body spectra if the emission originates from a range of disc semi-major
axes). 

\subsection{The maximum optical depth of a debris disc}
\label{subsect:fmax}

\begin{figure}
\begin{center}
\includegraphics[width=0.8\columnwidth]{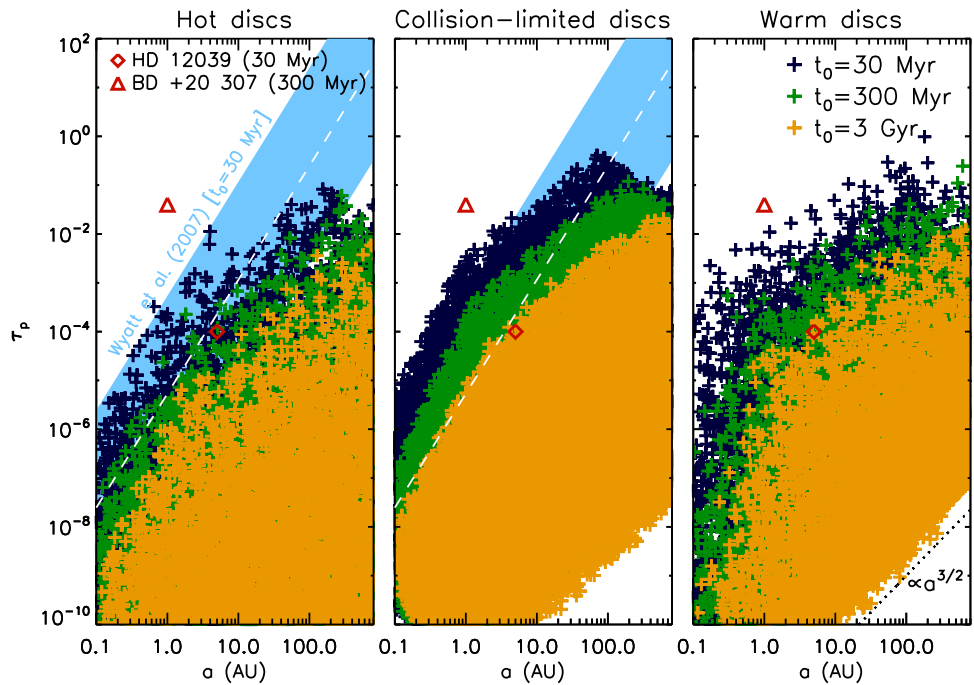}
\end{center}
\caption{Optical depth associated with dust grains in hot (left) and collision-limited (middle) planetesimal discs, as well as for planetesimals in warm discs (right).  Blue, green and yellow symbols are for planetesimal disc ages of 30 Myr, 300 Myr and 3 Gyr.  
  The white dashed line and light blue band represent the maximum optical depth and its
  associated uncertainty, respectively, as estimated by Wyatt et al.\ (2007)
  -- see equation (\ref{eq:wyatt_fmax}).
  Also shown are the observed values for HD 12039 and BD +20 307.}  
\label{fig:fmax}
\end{figure}

\begin{figure}
\begin{center}
\includegraphics[width=0.8\columnwidth]{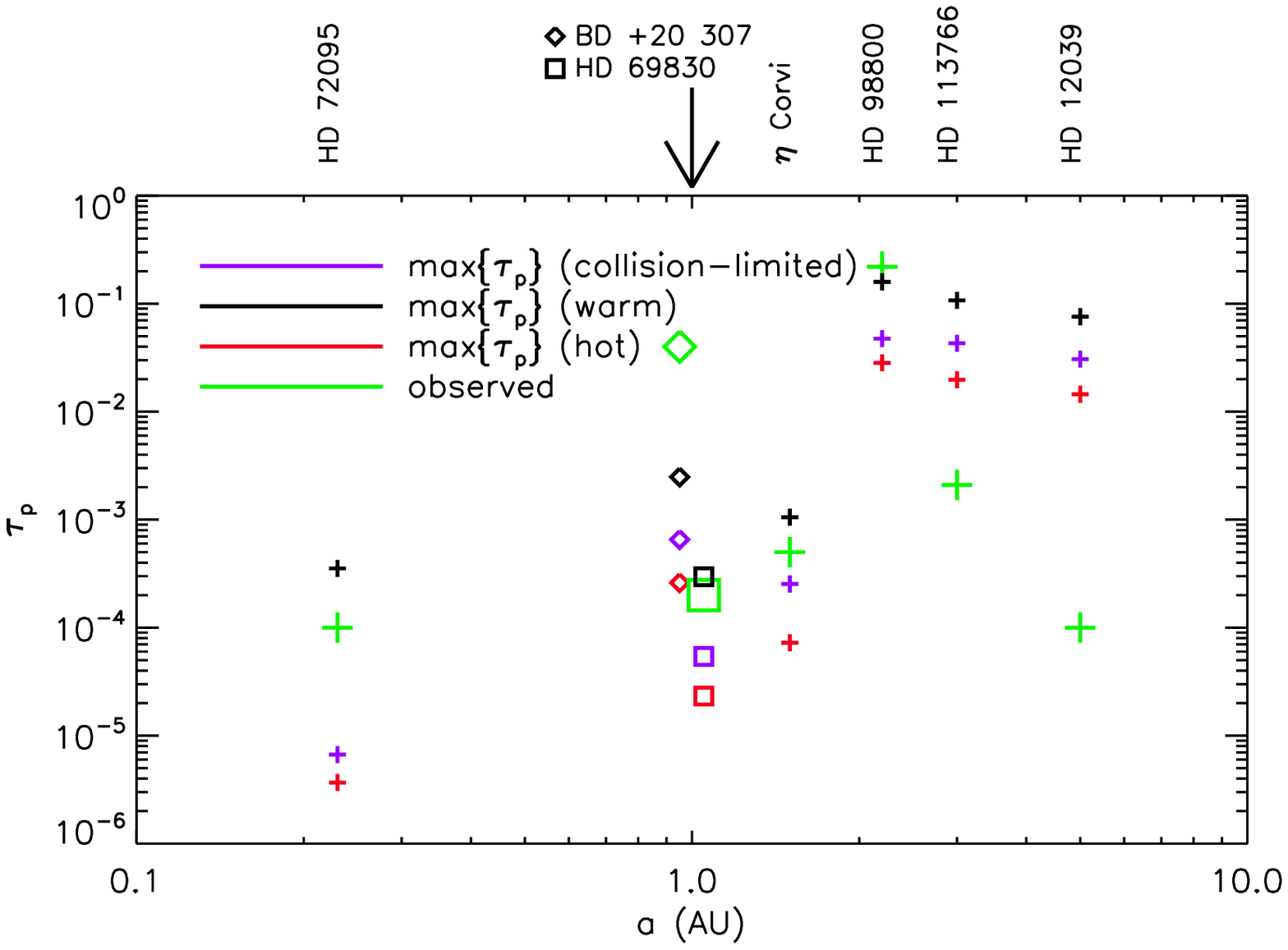}
\end{center}
\caption{Maximum optical depths for hot and warm discs as calculated by our
  model, compared to the observed values, for the 7 systems listed in Table 1
  of Wyatt et al. (2007).}  
\label{fig:transient}
\end{figure}

\cite{wy07a} argue that a simple model for the collisional evolution of
planetesimal discs implies that the maximum optical depth or fractional
bolometric disc luminosity is (their equation [21])
\be
\tau^{\rm (max)}_p = 1.6 \times 10^{-4} \left(\frac{a}{\mbox{AU}}\right)^{7/3} \left(\frac{t_0}{\mbox{Myr}} \right)^{-1}.
\label{eq:wyatt_fmax}
\ee
This result is based on several plausible but arbitrary assumptions
(planetesimal radius $r=2000\km$; rms eccentricity $e_0 = 0.05$, strength
$Q^\ast_D = 2 \times 10^6$ erg g$^{-1}$, etc). To examine the applicability of
this result, we employ the same Monte Carlo procedure used to produce Figure
\ref{fig:classify} to randomly generate hot, collision-limited, and warm planetesimal discs and
calculate the corresponding values of $\tau_p$.  For warm discs, the optical
depth is given by equation (\ref{eq:taupdef}) with $N$ and $r$ equal to the
number and radius of the planetesimals. For hot and collision-limited discs, the optical
depth is given by
\be
\tau_p={1\over 4a^2}\int Q_ar^2\,dN_{\rm dust};
\ee
taking the absorption efficiency $Q_a$ from equation (\ref{eq:mie}) and 
the number of particles $dN_{\rm dust}$ from equation (\ref{eq:dndustdr}) with
$q=7/2$, we find 
\begin{equation}
\tau_p = \frac{K}{a^2} \left(\frac{\lambda}{2\pi}\right)^{-1/2},
\end{equation} 
where $K$ is taken from equations (\ref{eq:kdef}) or (\ref{eq:kdefa}) for hot
and collision-limited discs, respectively. 
In the discussion below we assume $\lambda=70\,\mu$m. 

The results are shown in Figure \ref{fig:fmax} for disc
ages $t_0 = 30$ Myr, 300 Myr and 3 Gyr. We also show Wyatt et al.'s estimate
(\ref{eq:wyatt_fmax}) for $t_0=30$ Myr as a dashed white line; this is easily
scaled to other ages since $\tau^{\rm (max)}_p\propto 1/t_0$. Wyatt et al.\ estimate
the uncertainties involved to span $\sim 2$ orders of magnitude and this is
reflected in the light blue band shown in Figure \ref{fig:fmax}.

We are able to generate hot and collision-limited discs with optical depths substantially larger
than the estimate of equation (\ref{eq:wyatt_fmax}); however, these still lie
within the estimated range of uncertainty given by \cite{wy07a}.  A major
source of uncertainty is in the planetesimal strength $Q^\ast_D$. In the
calculations shown in Figure \ref{fig:fmax} we used a mass-dependent
$Q^\ast_D$ as defined in equation (\ref{eq:qast}). As a check, we carried out 
additional calculations assuming the constant value adopted by \cite{wy07a}
($Q^\ast_D=2 \times 10^6$ erg g$^{-1}$), and found that in this case our
results (not shown) agreed more closely with equation
(\ref{eq:wyatt_fmax}).

Our simulations show that the maximum optical depth is roughly $\propto
t^{-1}_0$, as predicted by equation (\ref{eq:wyatt_fmax}), but the scaling
with semi-major axis is quite different.  The numerical simulations of
\cite{loh08} also show that the scaling of $\tau_p$ with $a$ is generally more
complicated than a power law (see top right panel of their Figure 11).  

We also find that the maximum optical depth of warm discs can be almost an
order of magnitude higher than that of hot and collision-limited discs of the same age and
semi-major axis.  The absence of warm discs in the bottom right corner of
Figure \ref{fig:fmax} is simply a consequence of the defining condition of
warm discs, $t_c \lesssim t_0$ (equation [\ref{eq:tcolla}]), together with
equation (\ref{eq:tautcoll}) relating the collision time to the optical depth.
Discs exist below this line, but we label them `hot' rather than `warm'.  For collision-limited discs, 
there is a similar cut-off caused by the thin-disc condition (equation [\ref{eq:thin}]) and equation (\ref{eq:vcoll_limited}) imposing a maximum value for $m_{\rm max}$ (e.g., see Figure \ref{fig:sample_discs3}).

\cite{wy07a} point out that a number of debris discs with $a\lesssim 10\au$
have optical depths that exceed the limit (\ref{eq:wyatt_fmax}) by factors of
$10^3$ or more \citep[see also][]{mo09}. In Figure \ref{fig:fmax}, we show two
debris discs with small semi-major axes, HD 12039 ($t_0 = 30$ Myr) and BD +20
307 ($t_0 = 300$ Myr), taken from Table 1 of \cite{wy07a}.  These stars have a
range of spectral types, from F2 to K4, but our models based on a solar-type
host star should still be reasonably accurate. We verify that the optical
depth of BD +20 307 exceeds the maximum allowed for steady-state hot and collision-limited
planetesimal discs with an age equal to the stellar age, while the optical depth
of the disc around HD 12039 is consistent with steady-state models.

In Figure \ref{fig:transient} we show all seven debris discs listed in Table 1
of \cite{wy07a}.  For each system, we use the quoted values of the age $t_0$
and semi-major axis $a$ to compute the maximum value of $\tau_p$ for hot, collision-limited
and warm discs. The Figure shows that two systems (HD 113766 and HD 12039)
have optical depths consistent with a steady-state hot or collision-limited disc; one (BD +20 307)
has an optical depth that is inconsistent with a steady-state hot, collision-limited, or warm disc
(by factors of 100, 50 and 10, respectively); and four (HD 72095, HD 69830, $\eta$
Corvi, and HD 98800) are consistent with warm discs but not hot or collision-limited discs.
However, of these last four, the first three have $10\,\mu$m silicate features
in their spectra which imply that the IR emission comes from micron-sized
grains, thus ruling out warm discs as well. All of our conclusions about hot and collision-limited
discs are consistent with \cite{wy07a}, who suggest that the
dust arises from planetesimals that have been scattered to small semi-major
axes from a disc at much larger radii.

\subsection{Summary}

We have described a unified model of the evolution of gas-poor planetesimal
discs, which is general enough to apply to all Keplerian discs of solid
bodies, including debris discs, asteroid belts, and planetary systems. Our
model includes such processes as gravitational stability, evolution due to
dynamical chaos, gravitational scattering, radiation and stellar wind
pressure, Poynting--Robertson drag, and erosion or destruction by physical
collisions. We characterize the discs by four parameters: disc mass ($M_{\rm
  disc}$), disc semi-major axis ($a$), planetesimal size ($r$) and radial
velocity dispersion or rms eccentricity ($\sigma_r$ or $e_0$).  The salient
conclusions of our study include the following:

\begin{itemize}

\item Planetesimal discs can be categorized as dynamically `hot', `warm'
  or `cold' depending on whether the planetesimal orbits cross and therefore
  collide and whether the collisions are erosive/disruptive. In cold discs the
  orbits do not cross and collisions do not occur; in hot discs the orbits
  cross but the collision time is longer than the disc age, and in warm discs
  the collisions are frequent but gentle enough that they do not substantially
  erode the particles within the age of the disc.

\item Massive discs with small semi-major axes can only survive for Gyr
  timescales if they are cold. For example, after 3 Gyr hot discs at $1\au$ or
  $10\au$ cannot exceed $1.3\times10^{-4}M_\oplus$ or $1.5M_\oplus$ respectively
  (see \S\ref{sec:hotdisc}). Gravitational stability imposes an upper limit on
  the number of planetesimals per octave that can be present in a cold disc of
  given surface density; for example, a cold disc of mass $100M_\oplus$ cannot
  host more than 1--2 planetesimals per octave, while a disc of mass
  $1M_\oplus$ can host $\sim 10$ per octave (equation [\ref{eq:nmaxa}]).

\item Warm discs can survive for Gyr timescales over a wide range of
  semi-major axes and masses. At $1\au$ warm discs that survive for 3 Gyr must
  have mass $\lesssim 10^{-4}M_\oplus$; in this case the planetesimal radius
  is only 1 m, and warm discs composed of larger planetesimals must have even
  smaller masses (Figure \ref{fig:warm}). At larger semi-major axes the
  allowed masses of warm discs and the planetesimals within them are much
  larger (Figure \ref{fig:sample_discs3}). In some cases warm discs may be
  detectable from the IR emission from the planetesimals themselves.

\item Planetesimal discs can be detected by a wide variety of observational
  techniques, including transits, gravitational microlensing, radial-velocity
  variations, and `excess' IR emission (`debris discs'). With current
  technology the discs that can be detected by any of the first three methods
  are disjoint from those that can be detected in the IR (see Figure
  \ref{fig:classify}). Many possible long-lived planetesimal discs cannot be
  detected by any method at present. 

\end{itemize}

Despite the length of this paper, our analysis suffers from several
shortcomings. The assumption of a monodisperse planetesimal disc is
oversimplified, and probably incorrect given our limited understanding of disc
formation. We suspect that our results are reasonably accurate provided
that the total mass in the disc is dominated by planetesimals in a relatively
small mass range, but this suspicion should be tested by analysis of discs
with a range of planetesimal sizes. Our results also depend on a number of
poorly determined parameters of order unity (Table \ref{tab:nums}) and do not
incorporate a realistic model of the radial structure of the disc. 
In this paper we have deliberately ignored all considerations of the formation
process of planetesimal discs. It remains to be determined, by observations and
theory, which of the wide variety of possible long-lived planetesimal discs
are actually found in nature. 

\section*{Acknowledgments}

We acknowledge support from the Institute for Advanced Study, NASA
grant NNX08AH83G and NSF grant AST-0807444. We thank the anonymous referee for many thoughtful comments
that greatly improved the clarity and accuracy of our presentation.

\appendix

\section{Modifying various formulae in Dones \& Tremaine (1993)}
\label{append:dt93}

The rate of mass accretion in a rotating disc of planetesimals has been
evaluated by \cite{gl92} and Dones \& Tremaine (1993; hereafter DT).  We need
to modify their formulae, because their results are for the mass accretion
rate of a large body (a planet) on a circular orbit in the midplane of the
planetesimal disc, while we are interested in the rate for a typical
planetesimal in a monodisperse planetesimal disc.  Unless otherwise mentioned,
the notation used in this Appendix is the same as in the main text.

In the dispersion-dominated regime, the rate of mass accretion is given by
equations (72) and (90) of DT:
\be
\dot{M}=
\begin{cases}
2.7603 \Sigma \Omega R^2_p, & \Theta \ll 1,\\
6.0828 \Sigma \Omega^3 R_p R^3_{\rm H} \sigma^{-2}_r, & \Theta \gg 1,\\
\end{cases}
\ee 
where $R_p$ is the radius of the planet, $R_{\rm H} = a(M/M_\odot)^{1/3}$
is its Hill radius (as defined by DT, which is different from
the definition in the present paper) and $\sigma_r$ is the planetesimal
velocity dispersion in the radial direction.  In the shear-dominated regime,
we use equations (83) and (75) of DT:
\be \dot{M}=
\begin{cases}
10.1 \Sigma \Omega^2 R_p R^2_{\rm H} \sigma^{-1}_r, & \sigma_r \gtrsim \Omega
\sqrt{R_p R_{\rm H}},\\ 
6.47 \Sigma \Omega R^{1/2}_p R^{3/2}_{\rm H}, & \sigma_r \lesssim \Omega
\sqrt{R_p R_{\rm H}}.\\ 
\end{cases}
\ee

We first write these formulae in terms of the number density of planetesimals
in the midplane, $n_0$, using $\Sigma = \sqrt{2\pi} n_0 m \sigma_z/\Omega$
(see text below equation [18] of DT): 
\be
\dot{M}=
\begin{cases}
2.7603 \sqrt{2\pi} n_0 m \sigma_z R^2_p,\\
6.0828 \sqrt{2\pi} n_0 m \sigma_z \Omega^2 R_p R^3_{\rm H} \sigma^{-2}_r,\\
10.1 \sqrt{2\pi} n_0 m \sigma_z \Omega R_p R^2_{\rm H} \sigma^{-1}_r,\\
6.47 \sqrt{2\pi} n_0 m \sigma_z R^{1/2}_p R^{3/2}_{\rm H}.\\
\end{cases}
\ee

The collision time as defined in the present paper is $t^{-1}_c = \dot{M}/m$.  The following modifications are made:
\be
\begin{split}
&\sigma_{r,z} \rightarrow \sqrt{2} \sigma_{r,z},\\
&R_p \rightarrow 2r,\\
&R_{\rm H} = a\left(M/M_\odot\right)^{1/3} \rightarrow
a\left(2m/M_\odot\right)^{1/3}.\\ 
\end{split}
\ee 
The first modification comes from assuming the colliding bodies have the
same velocity dispersion, as opposed to one of them being on a circular orbit.
The second and third modifications arise both colliding bodies have the same
radius $r$and mass $m$, as opposed to one large body having radius $R$ and
mass $M$ while the other has negligible mass and radius. We also note that
$\sigma_z = \sigma_r (i_0/e_0)$, where we again choose $i_0/e_0=0.5$.  Thus,
the reciprocal of the collision time is 
\be 
t^{-1}_c =
\begin{cases}
2.7603 \times 4 \times \sqrt{\pi} n_0 r^2 \sigma_r,\\
6.0828 \times 2 \times \sqrt{\pi} n_0 \Omega^2 r a^3 \sigma^{-1}_r
\left(m/M_\odot\right),\\ 
10.1 \times 2^{7/6} \times \sqrt{\pi} n_0 \Omega r a^2 \left(m/M_\odot\right)^{2/3},\\
6.47 \times 2 \times \sqrt{\pi} n_0 r^{1/2} a^{3/2} \sigma_r
\left(m/M_\odot\right)^{1/2}.\\ 
\end{cases}
\ee

Finally, we decrease $t^{-1}_c$ by $\sqrt{2}$ since the vertical motions of
the particles imply that the mean density is reduced by this factor compared
to the midplane density.  We also replace $n_0$ by $\sqrt{2/\pi} {\cal N}
\Omega/\sigma_r$:
\be
t^{-1}_c = 
\begin{cases}
2.7603 \times 4 \times {\cal N} \Omega r^2,\\
6.0828 \times 4 \times {\cal N} \Omega r^2 \Theta,\\
10.1 \times 2^{7/6} \times {\cal N} \Omega^2 r a^2 \sigma^{-1}_r
\left(m/M_\odot\right)^{2/3},\\ 
6.47 \times 2 \times {\cal N} \Omega r^{1/2} a^{3/2} \left(m/M_\odot\right)^{1/2}.\\
\end{cases}
\ee

Comparing with equation (\ref{eq:tcolla}), we get $f_1 = 4 \times 2.7603/16 =
0.690$ and $f_2 = 6.0828/4 = 1.521$.  Similarly, by comparison with equation
(\ref{eq:tcollshear}), we get $f_4 = 2^{7/6} \times 10.1 = 22.67$ and $f_5 = 2
\times 6.47 = 12.94$. 

\bsp

\label{lastpage}

\end{document}